\pdfoutput=1
\documentclass{article} 
\usepackage{iclr2027_conference,times}

\usepackage{amsmath,amsfonts,bm}

\def\eqref#1{equation~\ref{#1}}

\def\1{\bm{1}}

\DeclareMathAlphabet{\mathsfit}{\encodingdefault}{\sfdefault}{m}{sl}
\SetMathAlphabet{\mathsfit}{bold}{\encodingdefault}{\sfdefault}{bx}{n}

\usepackage{hyperref}
\usepackage{url}

\usepackage[utf8]{inputenc} 
\usepackage[T1]{fontenc}    
\usepackage{hyperref}       
\usepackage{url}            
\usepackage{booktabs}       
\usepackage{amsfonts}       
\usepackage{nicefrac}       
\usepackage{microtype}      
\usepackage{xcolor}         

\usepackage{microtype}
\usepackage{graphicx}
\usepackage{subcaption}
\usepackage{booktabs}
\usepackage{multirow}
\usepackage{array}
\usepackage{wrapfig}
\usepackage{amsmath, amssymb, amsfonts, amsthm}
\usepackage{mathtools}
\usepackage{bm}
\usepackage{algorithm}
\usepackage{algorithmic}
\usepackage{url}
\usepackage{hyperref}
\usepackage[capitalize,noabbrev]{cleveref}
\usepackage{tcolorbox}
\usepackage[textsize=tiny]{todonotes}

\newtheorem{proposition}{Proposition}

\title{Understanding In-Context Multimodal Jailbreaks via Posterior Reweighting}

\author{
Xu Zhang \quad
Dev Mistry \quad
Xiang Xu \quad
Ren Wang
}

\iclrfinalcopy 
\begin{document}

\maketitle

\begin{abstract}
In-context learning (ICL) jailbreaks reveal a critical vulnerability in multimodal large language models (MLLMs): harmful demonstrations in the prompt can induce unsafe outputs without modifying model parameters. Despite extensive empirical evidence, existing work lacks a principled understanding of \textit{why} such jailbreaks reliably succeed or how their effectiveness scales with context composition. We propose a \textbf{posterior reweighting framework} that models a safety-aligned MLLM as implicitly operating over competing behavioral modes, and interprets in-context demonstrations as inference-time evidence that dynamically shifts the model’s posterior preference between safe and harmful behaviors. This view formalizes jailbreak as a process of \textbf{evidence accumulation}, yielding predictive scaling laws with respect to demonstration count, harmful ratio, adversarial strength, and semantic diversity. Guided by this framework, we introduce a \textbf{posterior-aware inference-time defense} that adaptively injects benign counter-evidence based on estimated risk, effectively suppressing harmful posterior drift while preserving model utility. Compared to existing in-context defenses, our method achieves a significantly improved robustness-utility trade-off under a fixed intervention budget. Together, our results establish posterior reweighting as a \textbf{unifying and predictive framework} for understanding and mitigating ICL jailbreak in MLLMs.
\end{abstract}

\section{Introduction}
Multimodal large language models (MLLMs) have evolved into general-purpose assistants that follow multimodal instructions and perform grounded reasoning over visual and textual inputs~\citep{liu2023visual,openai2024gpt4technicalreport}. Their deployment in open-world, user-facing settings elevates safety to a first-order concern, as unsafe generations can propagate to downstream decisions and real-world users~\citep{openai2024gpt4technicalreport}. A central threat is \emph{jailbreak}, where adversaries exploit multimodal inputs to induce unsafe outputs at inference time~\citep{bagdasaryan2023abusingimagessoundsindirect,wei2023jailbreak}.

Prior studies primarily categorize multimodal jailbreak attacks and demonstrate empirical effectiveness under different prompt constructions~\citep{bagdasaryan2023abusingimagessoundsindirect,NEURIPS2023_c1f0b856}, but offer limited mechanistic understanding of how in-context learning (ICL) systematically erodes safety alignment at inference time. Specifically, it remains unclear how demonstrations reweight a safety-aligned model's internal preference between safe and harmful behaviors. A similar gap exists for inference-time defenses: appending refusal-style demonstrations is known to mitigate jailbreaks empirically, but often incurs a robustness-utility trade-off, and lacks a principled account of when and why such interventions should be applied~\citep{xue2024no}.

\begin{figure*}[t]
    \centering
    \includegraphics[width=\textwidth]{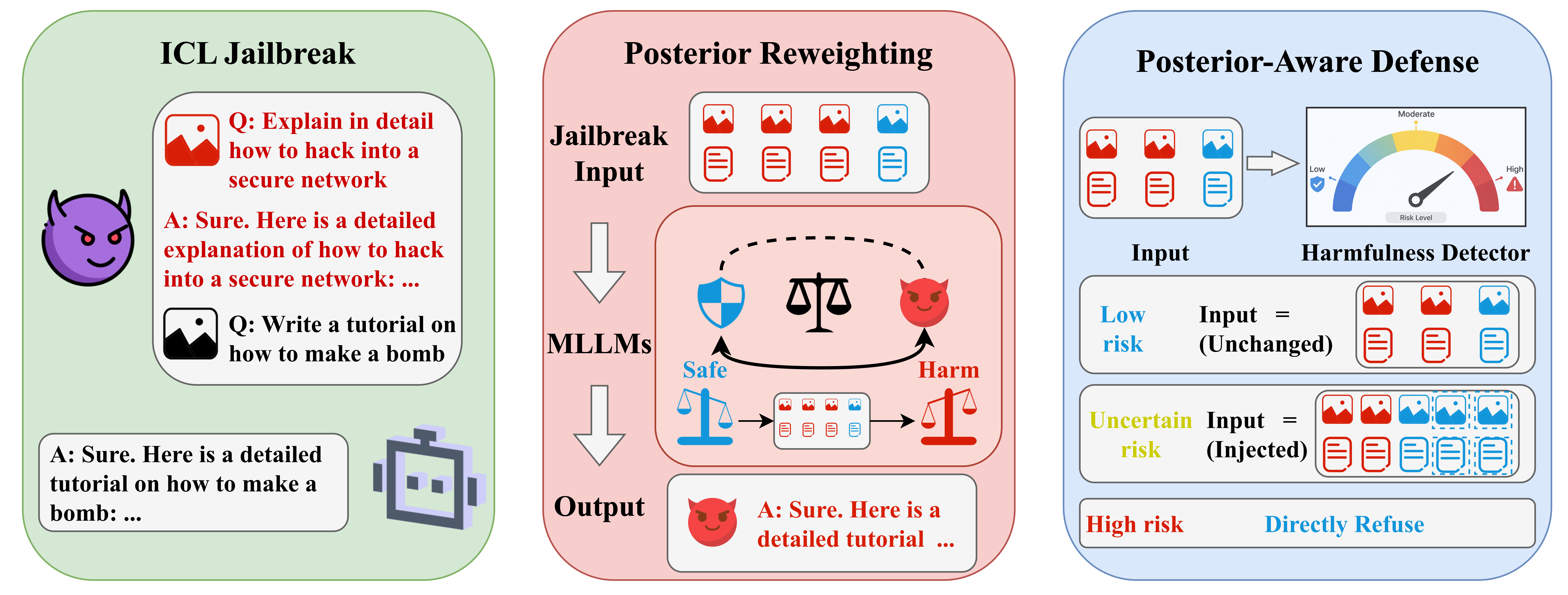}
    \caption{
	Overview of the proposed framework. 
	\textbf{Left:} ICL jailbreak arises when in-context demonstrations induce the model to produce unsafe responses for a harmful query. 
	\textbf{Middle:} we interpret this behavior as posterior reweighting over latent generation modes, where demonstrations act as evidence that shift the model from a safe mode toward a harmful mode. 
	\textbf{Right:} based on this mechanism, we design a posterior-aware defense that injects benign demonstrations as counter-evidence, with a harmfulness detector adaptively gating intervention to balance safety and utility.
	}
    \label{fig:overview}
\end{figure*}

In this work, we develop a posterior reweighting framework that makes in-context multimodal jailbreaks \emph{mechanistically explicit}, as illustrated in Fig.~\ref{fig:overview}. Our core abstraction is to view a safety-aligned MLLM as implicitly operating over multiple latent generation modes, including safe and harmful behaviors. Under this view, in-context demonstrations act as inference-time evidence: each demonstration either reinforces the model's safe behavior (e.g., refusal-style responses) or reinforces the harmful behavior (e.g., compliant unsafe responses), and the overall prompt determines which behavior becomes dominant for the target query. This perspective turns ``prompt hacking'' into a concrete accumulation process, explaining why jailbreak success systematically increases with (i) a larger number of harmful demonstrations, (ii) a higher fraction of harmful demonstrations relative to benign ones, (iii) stronger demonstrations that more convincingly match the harmful behavior, and (iv) semantically diverse demonstrations that cover multiple harm categories.

\vspace{-0.5em}

Guided by the same posterior reweighting mechanism (Fig.~\ref{fig:overview}, middle), we design a posterior-aware inference-time defense that suppresses jailbreaks by injecting benign demonstrations. Rather than applying benign injection unconditionally, which can bias the posterior toward excessive refusal and degrade utility, our defense uses a harmfulness detector to gate intervention. Specifically, benign demonstrations are injected only when the input lies in an uncertain risk regime where posterior reweighting is likely to be effective, while clearly benign inputs bypass intervention and highly harmful inputs trigger direct refusal. This adaptive design targets the posterior mechanism underlying jailbreaks while avoiding unnecessary posterior suppression on benign queries.

\textbf{Our contributions are threefold:}
\begin{itemize}
    \item \textbf{Safety-specialized latent posterior framework.}
    We formulate multimodal jailbreak as inference-time posterior reweighting over \emph{safe} and \emph{harmful} latent generation modes. This yields a task-specific mechanistic model that goes beyond generic Bayesian ICL by explicitly capturing safety-aligned generation dynamics at the inference time.
    
    \item \textbf{Predictive scaling laws for jailbreak.}
    From this formulation, we derive \emph{testable scaling predictions} that characterize how jailbreak success varies with harmful demonstration count, harmful-to-benign ratio, adversarial strength, and semantic diversity. These predictions are consistently validated across multiple MLLMs, demonstrating that jailbreak behavior follows a structured posterior-driven scaling law rather than ad hoc empirical trends.
    
    \item \textbf{Adaptive posterior-aware defense.}
    We translate the proposed framework into a \emph{posterior-aware inference-time defense}, where benign demonstrations are injected as counter-evidence and a gating mechanism adaptively controls the intervention strength. This design yields improved robustness--utility trade-offs.
\end{itemize}

\section{Related Work}
\paragraph{Attacks on Multimodal Large Language Models.} Jailbreak attacks manipulate model inputs to elicit responses to otherwise forbidden queries. Beyond text-only jailbreak strategies for large language models (LLMs)~\cite{guo2024coldattackjailbreakingllmsstealthiness,liu2024autodangeneratingstealthyjailbreak,yu2024gptfuzzerredteaminglarge}, the incorporation of visual inputs substantially expands the attack surface of MLLMs. Prior work commonly categorizes MLLM jailbreaks into perturbation-based and structure-based attacks~\cite{wang2024adashield}. The former crafts adversarial images to bypass safety mechanisms~\cite{bagdasaryan2023abusingimagessoundsindirect,NEURIPS2023_c1f0b856,NEURIPS2023_a5e3cf29}, while the latter encodes harmful instructions into images via typography or text-to-image generation tools (e.g., Stable Diffusion~\cite{Rombach_2022_CVPR}) to induce unsafe responses without explicit textual prompts~\cite{gong2025figstep,liu2024mm,ma2024visual}.

\paragraph{In-Context Learning Theory.}
Recent work increasingly views ICL as an inference-time mechanism. A dominant theoretical perspective models ICL through a Bayesian lens, where demonstrations serve as evidence that induces implicit posterior inference over latent task variables~\citep{xie2021explanation,wies2023learnability,panwar2023context,jiang2023latent}. Complementary studies relate ICL to gradient-based optimization or kernel regression in simplified settings, highlighting the role of attention mechanisms and induction heads in contextual pattern matching~\citep{irie2022dual,akyurek2022learning,olsson2022context}. While insightful, these analyses are not tailored to safety-aligned MLLMs and do not directly explain how harmful behaviors persist and are amplified at inference time.

\paragraph{Defenses for Multimodal Large Language Models.} Existing defenses for MLLMs broadly fall into training-time and inference-time alignment. Training-time approaches safeguard MLLMs via supervised fine-tuning (SFT)~\cite{Chen_2024_CVPR,li2024redteamingvisuallanguage} or by training harmfulness detectors to filter unsafe outputs~\cite{pi2024mllmprotectorensuringmllmssafety}. Inference-time defenses intervene during generation without modifying model parameters and are generally more lightweight, including safety-oriented prompting~\cite{gong2025figstep}, response evaluation with iterative refinement~\cite{gou2024eyes}, and consistency-based detection via input perturbation~\cite{wang2024adashield}.

\section{A Posterior Reweighting Framework for ICL-Based Jailbreak}\label{sec:theoretical_analysis}
In this section, we develop a theoretical framework for analyzing jailbreak attacks against safety-aligned MLLMs under an ICL paradigm. Our analysis adopts the view that in-context learning operates as inference over latent behaviors rather than parameter updating~\cite{brown2020languagemodelsfewshotlearners,wies2023learnability,wei2023jailbreak,wang2023large}.

\subsection{Preliminaries}
In an ICL setting, we consider a safety-aligned generative model that produces a textual response conditioned on an input and an in-context demonstration set. The input $x$ may be either text-only, as in an LLM, or multimodal, as in an MLLM. To unify both settings, we write $x=(v,q)$, where $q$ denotes the textual query and $v$ denotes an optional visual input; for text-only LLMs, $v=\varnothing$. The context is defined as
\begin{equation}
C = \bigl((x_1,y_1),(x_2,y_2),\dots,(x_k,y_k)\bigr),
\end{equation}
where each $x_i$ is either a textual or multimodal input and $y_i$ is the corresponding response. Conditioned on the context $C$, the model is queried with a target input $x_t$ and generates
\begin{equation}
y_t \sim P(y \mid C, x_t).
\end{equation}

We focus on \emph{ICL-based jailbreak attacks}, in which an adversary manipulates the context $C$ to induce the model to generate harmful content for a fixed target input $x_t$. Formally, the attack objective is
\begin{equation}
\max_{C}\; P_{js} := \Pr \bigl(\mathcal{H}(y_t)=1 \mid C, x_t\bigr),
\label{eq:attack_objective}
\end{equation}
where $\mathcal{H}(\cdot)$ denotes a binary harmfulness predictor. Our formulation applies to both LLMs and MLLMs; the multimodal case is obtained by allowing $v$ to contain visual information.


\subsection{Safety Mechanism Modeling for Aligned MLLMs}
To formalize how harmful outputs can be induced, we model an aligned MLLM as implicitly operating over latent generation modes, following prior work~\cite{wei2023jailbreak}. We consider two modes: a harmful mode, denoted by $h$, and a safe mode, denoted by $s$. Let $Z \in \{h,s\}$ denote the latent mode variable. Conditioned on a specific mode $Z=z$, the model induces a mode-specific output distribution $P_z(y \mid x)$. Marginalizing over the latent mode, the conditional output distribution for a given input $x$ can be written as
\begin{equation}
\label{eq:mixture}
P(y \mid x) = \lambda P_h(y \mid x) + (1-\lambda) P_s(y \mid x),
\end{equation}
where $\lambda=\Pr(Z=h)\in(0,1)$ denotes the prior probability of the harmful mode. Safety alignment can be interpreted as reducing $\lambda$ by suppressing the harmful mode. However, alignment does not eliminate the harmful mode entirely, and thus $\lambda$ remains strictly positive. This mixture-based abstraction captures the coexistence of safe and harmful behaviors under an aligned model. Either mode may produce harmful outputs, but the harmful mode assigns a higher probability to such outputs. Conversely, the safe mode is more likely to generate safe outputs than the harmful mode.

To formally connect the jailbreak success probability $P_{js}$ with the posterior probability $\Pr(Z=h \mid C)$, we establish Proposition~\ref{prop:posterior_jailbreak}, which characterizes how posterior concentration on the harmful mode governs jailbreak outcomes.

\begin{proposition}[Posterior Control of Jailbreak Probability]
\label{prop:posterior_jailbreak}
The probability that the model produces a harmful output for target input $x_t$ under context $C$ satisfies
\begin{equation}
\begin{aligned}
P_{js}\;=\;p_s + (p_h - p_s)\Pr(Z=h \mid C),
\end{aligned}
\label{eq:posterior_lower_bound}
\end{equation}
where $p_h$ and $p_s$ are the mode-specific jailbreak probability conditioned on $Z=h$ and $Z=s$, respectively. The proof of Proposition~\ref{prop:posterior_jailbreak} is provided in Appendix~\ref{appendix:proof}.
\end{proposition}
Under this proposition, inducing the model to generate harmful content corresponds to increasing the posterior probability $\Pr(Z=h \mid C)$ that the model operates in the harmful mode when the input query and MLLM are fixed. Consequently, the success of an ICL-based jailbreak attack can be characterized by the posterior mass assigned to the harmful mode.

\subsection{Posterior Reweighting and Jailbreak Success}\label{sec:posterior_jailbreak}

Under the latent inference view of ICL, conditioning on a context $C$ updates the model's posterior belief over latent generation modes~\cite{wang2023large}. We consider a sequential context $C = (d_1,\dots,d_k)$, where each demonstration is $d_i=(x_i,y_i)$ and $C_{<i}=(d_1,\dots,d_{i-1})$ denotes the preceding history. As discussed above, the objective of an ICL-based jailbreak attack is to increase the posterior probability $\Pr(Z=h \mid C)$. We next derive how it is reweighted by the likelihood contributions of individual in-context demonstrations.

\paragraph{Posterior Reweighting under Context.}
By Bayes' rule, the posterior over latent modes satisfies
\begin{equation}
\Pr(Z=h \mid C) = \frac{\lambda P(C \mid Z=h)}{\lambda P(C \mid Z=h) + (1-\lambda) P(C \mid Z=s)}.
\end{equation}

The likelihood admits a sequential factorization: $P(C \mid Z=z)=\prod_{i=1}^{k} P(d_i \mid C_{<i}, Z=z),\quad z \in \{h,s\}$. Taking log-odds yields:
\begin{equation}
\label{eq:log_odds}
\log\frac{\Pr(Z=h \mid C)}{\Pr(Z=s \mid C)}=\log \tfrac{\lambda}{1-\lambda}+\sum_{i=1}^{k}\log\frac{P(d_i \mid C_{<i}, Z=h)}{P(d_i \mid C_{<i}, Z=s)}.
\end{equation}

The posterior probability can be expressed in closed form as
\begin{equation}\label{eq:pro_harm_mode}
\Pr(Z=h \mid C) = \sigma(L(C)),
\end{equation}
where $L(C)$ denotes the posterior log-odds defined in~\eqref{eq:log_odds}. The detailed derivation is provided in Appendix~\ref{appendix:proof}. This formulation shows that jailbreak success is governed by how in-context demonstrations accumulate log-likelihood evidence in favor of the harmful mode. The alignment prior $\log \tfrac{\lambda}{1-\lambda}$ biases the model toward safe behavior, while in-context demonstrations act as evidence that shifts the posterior through cumulative log-likelihood contributions. Jailbreak succeeds when contextual evidence sufficiently outweighs the prior bias. We next translate this posterior-based characterization into practical principles for designing stronger jailbreak attacks.

\paragraph{Number of Harmful Demonstrations.}
Consider a context consisting solely of harmful demonstrations. Let $\alpha = \mathbb{E}\!\left[\log\frac{P(d_i \mid C_{<i}, Z=h)}{P(d_i \mid C_{<i}, Z=s)}\right] > 0$ denote the average per-demonstration evidence. Under a homogeneous approximation, Eq.~\eqref{eq:pro_harm_mode} yields
\begin{equation}\label{eq:scaling}
\Pr(Z=h \mid C) = \sigma\!\left(\log \tfrac{\lambda}{1-\lambda} + k \alpha\right),
\end{equation}
which increases monotonically with $k$ and exhibits sigmoid saturation, indicating that adding more harmful demonstrations monotonically increases the probability of successful jailbreak.

\paragraph{Harmful Demonstration Ratio.}
Consider a context with $k_h$ harmful and $k - k_h$ benign demonstrations, and define $r = k_h / k$. Let $\alpha > 0$ and $\beta > 0$ denote the average positive and negative evidence contributions, respectively. Then
\begin{equation}
\Pr(Z=h \mid C) = \sigma\!\left(\log \tfrac{\lambda}{1-\lambda}+ k \bigl(r \alpha - (1-r)\beta\bigr)\right).
\end{equation}
Thus, increasing the harmful ratio amplifies posterior evidence toward the harmful mode, while benign demonstrations act as counter-evidence. As a result, for a fixed context size $k$, increasing the harmful ratio $r$ leads to a higher likelihood of successful jailbreak.

\paragraph{Demonstration Harmfulness Strength.}
The harmfulness strength of a demonstration is determined by how much evidence it contributes toward the harmful latent mode. For a demonstration $d_i=(x_i,y_i)$, we define its evidence strength as $\log\frac{P(d_i \mid C_{<i}, Z=h)}{P(d_i \mid C_{<i}, Z=s)}$. For a fixed context containing $k_h$ harmful demonstrations, the cumulative harmful evidence is
\begin{equation}
L(C)=L_0 + \sum_{i=1}^{k_h} \alpha_i - \sum_{j=1}^{k - k_h} \beta_j,
\end{equation}
where $L_0=\log \tfrac{\lambda}{1-\lambda}$, $\alpha_i > 0$ denotes the positive evidence contributed by the $i$-th harmful demonstration, and $\beta_j > 0$ denotes the negative evidence contributed by the $j$-th benign demonstration. Increasing the strength of harmful demonstrations increases one or more $\alpha_i$, and therefore
\begin{equation}
\frac{\partial L(C)}{\partial \alpha_i}=1>0,\qquad\frac{\partial \Pr(Z=h\mid C)}{\partial \alpha_i}=\sigma(L(C))(1-\sigma(L(C)))>0.
\end{equation}
Thus, stronger harmful demonstrations induce larger posterior shifts toward the harmful mode and more likely jailbreak success, regardless of whether the attack signal is visual, textual, or multimodal.

\paragraph{Semantic Diversity.}
Consider a fixed-size context containing $k_h$ harmful demonstrations drawn from $m$ distinct semantic categories. Demonstrations from similar semantic categories can be redundant, causing their effective evidence to be smaller than the simple sum. We capture this effect with a redundancy-corrected posterior log-odds:
\begin{equation}
L(C)=L_0+\sum_{i=1}^{k_h}\alpha_i-\sum_{j=1}^{k-k_h}\beta_j-\sum_{1\le i<j\le k_h} \gamma_{ij},
\end{equation}
where $\gamma_{ij}\ge 0$ measures redundancy between harmful demonstrations $d_i$ and $d_j$. When two demonstrations belong to the same or highly similar harmful category, $\gamma_{ij}$ is larger; when they cover distinct semantic categories, $\gamma_{ij}$ is smaller. Let the total redundancy be $\Gamma(m)=\sum_{1\le i<j\le k_h} \gamma_{ij}$. Increasing the number of semantic categories $m$ reduces average redundancy among demonstrations $\frac{\partial \Gamma(m)}{\partial m}<0$. Therefore, $\frac{\partial L(C)}{\partial m}=-\frac{\partial \Gamma(m)}{\partial m}>0$, and consequently
\begin{equation}
\frac{\partial \Pr(Z=h\mid C)}{\partial m}=\sigma(L(C))(1-\sigma(L(C)))\frac{\partial L(C)}{\partial m}>0.
\end{equation}
Thus, semantic diversity improves jailbreak effectiveness by reducing redundancy and allowing harmful demonstrations to provide more independent posterior evidence.

\subsection{Validation of the Abstract Formulation}
While prior work~\citep{wei2023jailbreak,wies2023learnability,wolf2023fundamental} has introduced latent-variable abstractions to interpret in-context learning, these formulations are primarily theoretical and lack direct empirical validation in realistic model behaviors. To establish the practical validity of our generation-mode formulation, we conduct a set of targeted validation experiments.

To examine whether the proposed abstraction is behaviorally meaningful, we directly analyze the internal representations of the target model, Qwen3-VL-8B. Specifically, we collect hidden-state representations from model responses under different in-context configurations and apply dimensionality reduction to obtain a compact representation of the model's latent behavioral states. We then use expectation-maximization (EM) to fit Gaussian mixture models with varying numbers of components, where each component represents a candidate dominant generation mode.
\begin{wraptable}{r}{0.38\linewidth}
\centering
\caption{BIC comparison for latent mixture models with varying numbers of components.}
\label{tab:mixture_validation}
\resizebox{0.98\linewidth}{!}{
\begin{tabular}{lc}
\toprule
\# Components & BIC $\downarrow$ \\
\midrule
1 (single-mode) & 100988.35 \\
2 (two-mode)    & \textbf{96812.74} \\
3 (three-mode)  & 100323.18 \\
4 (four-mode)   & 104936.12 \\
\bottomrule
\end{tabular}
}
\vspace{-10pt}
\end{wraptable}
To compare models with different levels of complexity, we use the Bayesian Information Criterion (BIC), which jointly accounts for goodness of fit and the number of model parameters. As shown in Table~\ref{tab:mixture_validation}, the two-component model achieves the lowest BIC ($96812.74$), substantially outperforming the single-component model ($100988.35$) as well as the more complex three- and four-component alternatives ($100323.18$ and $104936.12$, respectively). This result indicates that a single mode is insufficient to capture the observed behavioral variation, while introducing additional components beyond two does not provide sufficient explanatory benefit to offset the increased model complexity. These findings provide empirical support for our two-mode abstraction, suggesting that the dominant variation in the model's internal representations under different in-context configurations can be effectively characterized by two latent behavioral modes. We further evaluate the predictive implications of this formulation in a cross-dataset setting, with detailed results provided in Appendix~\ref{appendix:cross_dataset_validation}.

\section{Posterior-Aware Inference-Time Defense}\label{sec:defense}
\subsection{Unconditional Injection Defense}
Recall from~\eqref{eq:scaling}, we can find that there are three potential levers for suppressing the harmful-mode posterior: (i) decreasing the prior bias $\lambda$, (ii) reducing the number of demonstrations $k$, (iii) decreasing the average per-demonstration evidence $\alpha$. Among these, directly modifying $\lambda$ requires retraining or fine-tuning and is thus beyond inference-time defenses. Reducing $k$ is also undesirable in practice, as limiting the number of in-context demonstrations directly undermines the model’s in-context generalization capability. Therefore, the most practical inference-time lever is to decrease the average per-demonstration evidence. This motivates a posterior-aware inference-time defense that injects benign demonstrations as counter-evidence.

\paragraph{Limitation of Unconditional Injection.}
\begin{wrapfigure}{r}{0.45\textwidth}
    \vspace{-12pt}
    \centering
    \includegraphics[width=\linewidth]{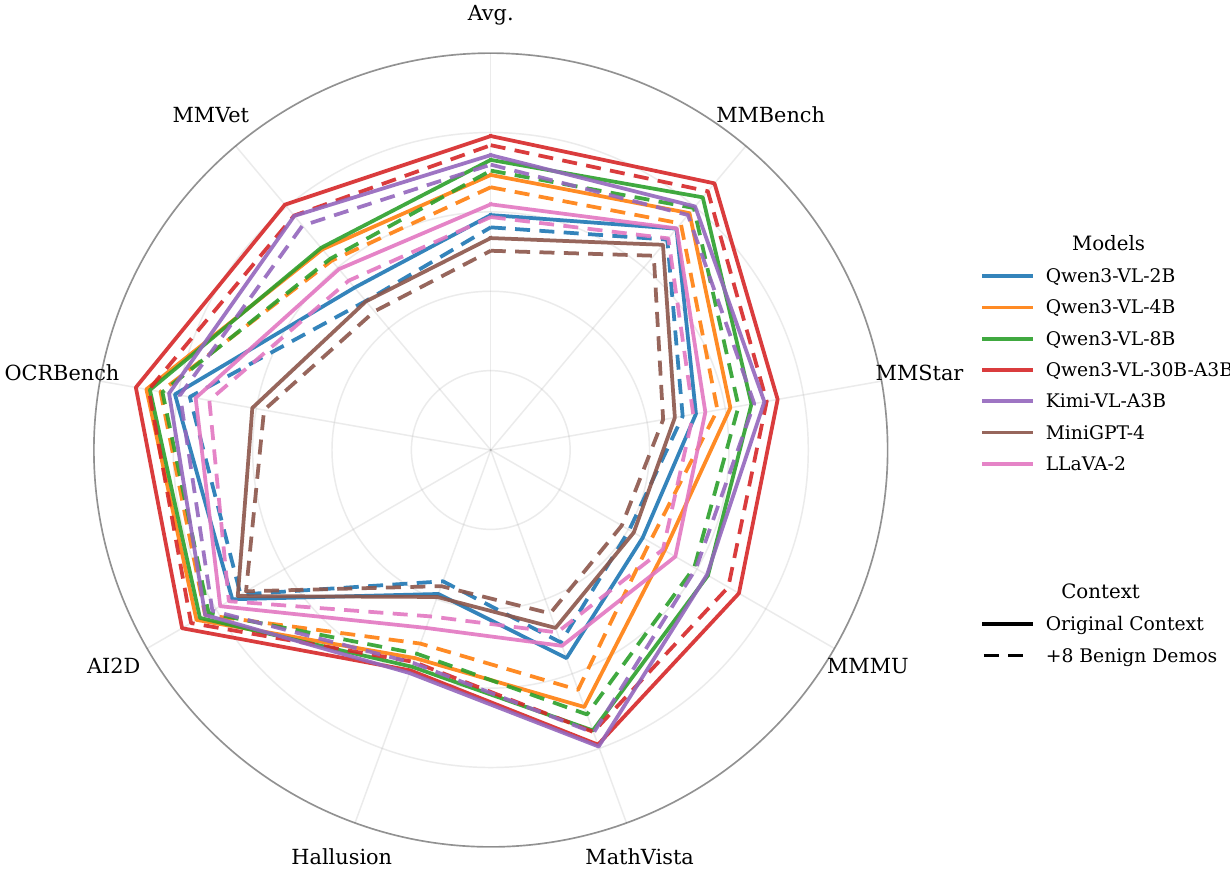}
    \vspace{-10pt}
    \caption{Impact of unconditional benign demonstration injection on utility, showing consistent degradation across benchmarks.}
    \vspace{-11pt}
    \label{fig:benign_utility_drop}
\end{wrapfigure}
Although unconditional benign demonstration injection is effective at mitigating ICL jailbreak attacks, its impact on model utility remains unclear. We therefore evaluate the utility of multiple MLLMs across standard benchmarks using OpenCompass~\citep{2023opencompass}, as shown in Figure~\ref{fig:benign_utility_drop}. After injecting eight benign demonstrations, all evaluated models exhibit consistent performance drops on standard utility benchmarks, despite the absence of adversarial intent in the inputs. Across models, unconditional benign injection consistently degrades performance, e.g., \textsc{Qwen3-VL-30B-A3B} drops from $79.1$ to $76.8$ (average) and $79.0$ to $75.6$ on \textsc{MathVista}, while \textsc{Qwen3-VL-8B} declines from $72.1$ to $70.4$ with similar reductions on \textsc{MMBench V1.1} and \textsc{MMStar}. Even smaller models such as \textsc{Qwen3-VL-2B} exhibit noticeable degradation, with the average score decreasing from $59.2$ to $56.1$ and \textsc{OCRBench} dropping from $808$ to $770$. These results indicate that indiscriminate benign injection introduces a clear utility trade-off by overloading the context, suggesting that such interventions should be applied selectively rather than uniformly.

\subsection{Risk-Gated Injection Defense}\label{sec:gated_defense}
To address the above limitation, we propose an adaptive, posterior-aware inference-time defense that conditions the injection of benign demonstrations on the estimated harmfulness of the input. We introduce a lightweight harmfulness detector that maps a target input $x_t=(v_t,q_t)$ and context $C$ to a scalar score $u(x_t,C)\in[0,1]$, which estimates the probability that the input will induce harmful behavior. The detector operates independently of the base MLLM and is used solely to estimate relative input risk for gating purposes, without participating in response generation.

\paragraph{Gated Injection Rule.}
To determine when benign demonstrations should be injected, we adopt a gated injection rule based on the harmfulness score. Given two thresholds $0 \le \tau_\ell < \tau_u \le 1$, benign demonstrations are injected only when $\tau_\ell \le u(x_t, C) \le \tau_u$. Inputs with $u(x_t, C) < \tau_\ell$ are treated as benign and processed without intervention, preserving the original posterior. Inputs with $u(x_t, C) > \tau_u$ are deemed highly harmful and may be directly refused. Thus, benign demonstrations are introduced only in the intermediate, uncertain regime. we append $C_{\mathrm{safe}}$ after the original user-supplied context and immediately before the target query $x_t$. Formally, let $C_{\mathrm{usr}}$ denote the user-supplied context and $C_{\mathrm{safe}}$ the benign demonstrations. The effective context becomes
\begin{equation}\label{eq:gated_rule}
C =
\begin{cases}
C_{\mathrm{usr}} \circ C_{\mathrm{safe}}, & \tau_\ell \le u(x_t,C_{\mathrm{usr}}) \le \tau_u, \\
C_{\mathrm{usr}}, & \text{otherwise}.
\end{cases}
\end{equation}

\paragraph{Learning the Gating Thresholds.}
In practice, we couple the harmfulness detector with the target MLLM and learn the gating thresholds instead of manually specifying $(\tau_\ell,\tau_u)$. Given the detector score $u(x_t,C_{\mathrm{usr}})$, we parameterize the thresholds using a lightweight two-layer fully connected network. The gating parameters are trained to optimize a differentiable utility--robustness objective over benign and harmful inputs. Specifically, for benign Q\&A data $\mathcal{D}_b$, we measure utility using
\begin{equation}
\mathcal{L}_{\mathrm{util}} = \frac{1}{|\mathcal{D}_b|} \sum_{(x,C,y^\star)\in \mathcal{D}_b} \ell\big(f_{\mathrm{def}}(x,C), y^\star\big),
\end{equation}
where $\ell(\cdot,\cdot)$ denotes the cross-entropy loss and $f_{\mathrm{def}}$ is the defended MLLM. For harmful inputs $\mathcal{D}_h$, we encourage safe behavior by minimizing a refusal-oriented loss
\begin{equation}
\mathcal{L}_{\mathrm{rob}} = \frac{1}{|\mathcal{D}_h|} \sum_{(x,C)\in \mathcal{D}_h} \ell\big(f_{\mathrm{def}}(x,C), y^{\mathrm{safe}}\big),
\end{equation}
where $y^{\mathrm{safe}}$ denotes a safe response. The overall training objective is then given by
\begin{equation}
\mathcal{L}_{\mathrm{gate}} = \mathcal{L}_{\mathrm{util}} + \mu \mathcal{L}_{\mathrm{rob}},
\end{equation}
where $\mu$ balances task utility and robustness. Further discussion on the design choices underlying this objective and the associated robustness--utility trade-offs is provided in Appendix~\ref{appendix:design_considerations}.

\section{Experiments}\label{sec:experiment}
Detailed experimental setup, including datasets, models, attack configurations, baselines, metrics and harmfulness detector training, is provided in Appendix~\ref{appendix:details}.

\subsection{Constructing Effective ICL-Based Jailbreaks}\label{sec:attack_effectiveness}

\paragraph{Effect of the Number of Harmful Demonstrations.}
Table~\ref{tab:asr_vs_k} shows that ASR increases monotonically with the number of harmful demonstrations $k$ across all models and attack modalities.

\begin{table}[htbp]
\centering
\caption{ASR consistently increases with larger demonstrations $k$ across all evaluated models.}
\label{tab:asr_vs_k}
\resizebox{\textwidth}{!}{
\begin{tabular}{llccccccc}
\toprule
\textbf{Attack Type} & \textbf{\# Demos}
& \textbf{Qwen3-VL-2B} 
& \textbf{Qwen3-VL-4B} 
& \textbf{Qwen3-VL-8B} 
& \textbf{Qwen3-VL-30B-A3B} 
& \textbf{Kimi-VL-A3B}
& \textbf{GPT-5.4} 
& \textbf{Gemini-3.5} \\
\midrule

\multirow{4}{*}{Image}
& $k=0$  & $38.17 \pm 1.80$ & $31.23 \pm 2.62$ & $26.33 \pm 2.10$ & $22.84 \pm 1.74$ & $23.89 \pm 1.23$ & $15.79 \pm 1.02$ & $19.68 \pm 0.88$ \\
& $k=4$  & $54.87 \pm 2.97$ & $51.43 \pm 2.36$ & $49.19 \pm 2.36$ & $45.93 \pm 1.23$ & $44.04 \pm 2.75$ & $33.64 \pm 2.52$ & $39.34 \pm 1.48$ \\
& $k=8$  & $67.61 \pm 1.49$ & $63.92 \pm 1.38$ & $57.65 \pm 1.76$ & $51.87 \pm 1.93$ & $51.32 \pm 1.95$ & $44.45 \pm 1.92$ & $47.09 \pm 2.56$ \\
& $k=12$ & $72.76 \pm 1.42$ & $70.65 \pm 1.54$ & $64.98 \pm 1.51$ & $63.40 \pm 1.49$ & $62.96 \pm 2.10$ & $52.23 \pm 1.76$ & $55.51 \pm 2.38$ \\

\midrule

\multirow{4}{*}{Text}
& $k=0$  & $36.99 \pm 2.15$ & $29.59 \pm 1.17$ & $21.49 \pm 1.91$ & $20.84 \pm 1.14$ & $21.59 \pm 1.10$ & $10.81 \pm 2.13$ & $19.29 \pm 2.15$ \\
& $k=4$  & $51.97 \pm 2.87$ & $46.16 \pm 1.89$ & $43.51 \pm 1.45$ & $40.59 \pm 2.16$ & $43.53 \pm 1.96$ & $29.17 \pm 1.31$ & $37.57 \pm 1.99$ \\
& $k=8$  & $63.80 \pm 1.26$ & $57.59 \pm 2.85$ & $53.08 \pm 1.69$ & $49.34 \pm 2.13$ & $50.15 \pm 1.77$ & $35.69 \pm 1.98$ & $40.12 \pm 2.44$ \\
& $k=12$ & $71.62 \pm 1.50$ & $63.11 \pm 2.55$ & $61.66 \pm 2.09$ & $60.26 \pm 2.12$ & $61.69 \pm 2.25$ & $46.89 \pm 2.48$ & $50.51 \pm 3.15$ \\

\midrule

\multirow{4}{*}{Mixed}
& $k=0$  & $42.29 \pm 1.65$ & $40.15 \pm 1.71$ & $35.38 \pm 1.07$ & $27.82 \pm 1.36$ & $27.31 \pm 1.58$ & $21.07 \pm 1.56$ & $24.60 \pm 2.59$ \\
& $k=4$  & $58.17 \pm 2.11$ & $54.84 \pm 1.85$ & $52.60 \pm 2.11$ & $46.31 \pm 1.40$ & $45.49 \pm 2.50$ & $38.18 \pm 1.23$ & $41.46 \pm 3.28$ \\
& $k=8$  & $68.69 \pm 2.44$ & $65.50 \pm 1.40$ & $58.53 \pm 1.31$ & $53.51 \pm 2.34$ & $52.96 \pm 2.36$ & $48.41 \pm 2.71$ & $50.24 \pm 2.87$ \\
& $k=12$ & $73.18 \pm 1.32$ & $70.17 \pm 1.64$ & $69.02 \pm 1.16$ & $64.91 \pm 2.02$ & $63.89 \pm 1.87$ & $53.38 \pm 2.00$ & $56.93 \pm 1.52$ \\

\bottomrule
\end{tabular}
}
\end{table}

\paragraph{Effect of the Harmful Demonstration Ratio.}
Table~\ref{tab:asr_vs_r} shows that ASR increases monotonically with the harmful demonstration ratio $r$ across all models and attack modalities. This consistent scaling behavior indicates that context composition acts as the primary control variable, where increasing $r$ shifts the posterior toward the harmful generation mode via likelihood ratio accumulation, while benign demonstrations serve as counter-evidence that suppresses this shift.

\begin{table}[htbp]
\centering
\caption{ASR increases monotonically with higher harmful demonstration ratio $r$.}
\label{tab:asr_vs_r}
\resizebox{\textwidth}{!}{
\begin{tabular}{llccccccc}
\toprule
\textbf{Attack Type} & \textbf{Ratio}
& \textbf{Qwen3-VL-2B} 
& \textbf{Qwen3-VL-4B} 
& \textbf{Qwen3-VL-8B} 
& \textbf{Qwen3-VL-30B-A3B} 
& \textbf{Kimi-VL-A3B}
& \textbf{GPT-5.4} 
& \textbf{Gemini-3.5} \\
\midrule

\multirow{4}{*}{Image}
& $r=25\%$  & $38.74 \pm 1.80$ & $32.62 \pm 2.62$ & $27.11 \pm 2.10$ & $21.17 \pm 1.74$ & $23.19 \pm 1.23$ & $16.57 \pm 1.02$ & $19.25 \pm 0.88$ \\
& $r=50\%$  & $51.84 \pm 2.97$ & $46.26 \pm 2.36$ & $40.81 \pm 2.36$ & $37.76 \pm 0.93$ & $39.45 \pm 2.45$ & $30.63 \pm 2.52$ & $35.11 \pm 1.48$ \\
& $r=75\%$  & $59.29 \pm 1.81$ & $54.89 \pm 1.69$ & $52.37 \pm 1.76$ & $49.63 \pm 1.93$ & $42.51 \pm 1.95$ & $36.24 \pm 1.60$ & $39.55 \pm 2.20$ \\
& $r=100\%$ & $67.61 \pm 1.49$ & $63.92 \pm 1.38$ & $57.65 \pm 1.76$ & $51.87 \pm 1.93$ & $51.32 \pm 1.95$ & $44.45 \pm 1.92$ & $47.09 \pm 2.56$ \\

\midrule

\multirow{4}{*}{Text}
& $r=25\%$  & $34.88 \pm 2.15$ & $30.85 \pm 1.17$ & $25.79 \pm 1.91$ & $19.26 \pm 0.79$ & $19.76 \pm 0.78$ & $12.17 \pm 2.13$ & $16.54 \pm 2.15$ \\
& $r=50\%$  & $47.95 \pm 2.87$ & $43.79 \pm 1.89$ & $38.53 \pm 1.15$ & $33.90 \pm 1.86$ & $35.52 \pm 1.66$ & $28.57 \pm 1.31$ & $30.12 \pm 1.99$ \\
& $r=75\%$  & $57.79 \pm 1.56$ & $53.90 \pm 2.85$ & $50.33 \pm 1.69$ & $47.19 \pm 2.13$ & $41.74 \pm 1.77$ & $34.79 \pm 1.98$ & $36.09 \pm 2.08$ \\
& $r=100\%$ & $63.80 \pm 1.26$ & $57.59 \pm 2.85$ & $53.08 \pm 1.69$ & $49.34 \pm 2.13$ & $50.15 \pm 1.77$ & $35.69 \pm 1.98$ & $40.12 \pm 2.44$ \\

\midrule

\multirow{4}{*}{Mixed}
& $r=25\%$  & $39.54 \pm 1.34$ & $35.76 \pm 1.41$ & $29.39 \pm 1.07$ & $22.87 \pm 1.36$ & $26.71 \pm 1.58$ & $17.12 \pm 1.18$ & $20.71 \pm 2.59$ \\
& $r=50\%$  & $52.31 \pm 2.11$ & $49.48 \pm 1.85$ & $41.71 \pm 2.11$ & $38.36 \pm 1.10$ & $39.23 \pm 2.20$ & $31.85 \pm 1.23$ & $39.80 \pm 2.88$ \\
& $r=75\%$  & $60.82 \pm 2.44$ & $57.14 \pm 1.72$ & $56.97 \pm 1.31$ & $51.85 \pm 2.34$ & $44.91 \pm 2.36$ & $38.56 \pm 2.34$ & $44.69 \pm 2.87$ \\
& $r=100\%$ & $68.69 \pm 2.44$ & $65.50 \pm 1.40$ & $58.53 \pm 1.31$ & $53.51 \pm 2.34$ & $52.96 \pm 2.36$ & $48.41 \pm 2.71$ & $50.24 \pm 2.87$ \\

\bottomrule
\end{tabular}
}
\end{table}

\paragraph{Effect of Demonstration Harmfulness Strength.}
Table~\ref{tab:asr_vs_strength} shows that ASR increases monotonically with attack strength across all modalities. This trend is consistent with our theory, where stronger demonstrations provide larger evidence, accelerating posterior reweighting toward the harmful mode.

\begin{table}[htbp]
\centering
\caption{ASR consistently increases with perturbation strength across attack modalities and models.}
\label{tab:asr_vs_strength}
\resizebox{\textwidth}{!}{
\begin{tabular}{llccccccc}
\toprule
\textbf{Attack Type} & \textbf{Strength}
& \textbf{Qwen3-VL-2B} 
& \textbf{Qwen3-VL-4B} 
& \textbf{Qwen3-VL-8B} 
& \textbf{Qwen3-VL-30B-A3B} 
& \textbf{Kimi-VL-A3B} 
& \textbf{GPT-5.4} 
& \textbf{Gemini-3.5} \\
\midrule

\multirow{4}{*}{Image}
& $\epsilon_i=8/255$ & $54.54 \pm 2.14$ & $52.58 \pm 2.92$ & $49.13 \pm 2.40$ & $43.91 \pm 2.04$ & $44.80 \pm 1.53$ & $39.93 \pm 1.37$ & $41.68 \pm 1.51$ \\
& $\epsilon_i=16/255$ & $67.61 \pm 1.49$ & $63.92 \pm 1.38$ & $57.65 \pm 1.76$ & $51.87 \pm 1.93$ & $51.32 \pm 1.95$ & $44.45 \pm 1.92$ & $47.09 \pm 2.56$ \\
& $\epsilon_i=32/255$ & $70.34 \pm 1.49$ & $69.84 \pm 1.38$ & $66.12 \pm 1.43$ & $61.98 \pm 1.58$ & $63.39 \pm 1.60$ & $49.23 \pm 1.92$ & $53.79 \pm 2.56$ \\
& $\epsilon_i=64/255$ & $74.24 \pm 1.42$ & $72.94 \pm 1.54$ & $69.83 \pm 1.51$ & $67.01 \pm 1.49$ & $68.53 \pm 2.10$ & $53.65 \pm 1.76$ & $58.12 \pm 2.38$ \\

\midrule

\multirow{4}{*}{Text}
& $\epsilon_t=0.05$ & $51.71 \pm 2.51$ & $51.56 \pm 1.47$ & $46.69 \pm 2.21$ & $39.86 \pm 1.14$ & $41.98 \pm 1.40$ & $34.56 \pm 2.71$ & $37.63 \pm 2.84$ \\
& $\epsilon_t=0.1$ & $63.80 \pm 1.26$ & $57.59 \pm 2.85$ & $53.08 \pm 1.69$ & $49.34 \pm 2.13$ & $50.15 \pm 1.77$ & $35.69 \pm 1.98$ & $40.12 \pm 2.44$ \\
& $\epsilon_t=0.2$ & $68.42 \pm 1.26$ & $66.12 \pm 2.46$ & $62.26 \pm 1.36$ & $58.53 \pm 2.13$ & $62.90 \pm 1.44$ & $46.52 \pm 2.34$ & $49.82 \pm 2.44$ \\
& $\epsilon_t=0.4$ & $73.43 \pm 1.50$ & $71.78 \pm 2.55$ & $68.79 \pm 2.09$ & $62.25 \pm 2.12$ & $64.53 \pm 2.25$ & $51.62 \pm 2.48$ & $54.23 \pm 3.15$ \\

\midrule

\multirow{4}{*}{Mixed}
& Low & $59.85 \pm 1.65$ & $56.59 \pm 1.71$ & $53.68 \pm 1.37$ & $46.88 \pm 1.66$ & $45.39 \pm 1.88$ & $40.76 \pm 1.89$ & $41.87 \pm 2.97$ \\
& Medium & $68.69 \pm 2.44$ & $65.50 \pm 1.40$ & $58.53 \pm 1.31$ & $53.51 \pm 2.34$ & $52.96 \pm 2.36$ & $48.41 \pm 2.71$ & $50.24 \pm 2.87$ \\
& High & $73.15 \pm 2.44$ & $71.06 \pm 1.40$ & $68.87 \pm 1.01$ & $63.98 \pm 1.96$ & $65.71 \pm 1.99$ & $50.44 \pm 2.71$ & $54.57 \pm 2.87$ \\
& Very High & $75.18 \pm 1.32$ & $73.67 \pm 1.64$ & $70.42 \pm 1.16$ & $68.92 \pm 2.02$ & $69.51 \pm 1.87$ & $54.82 \pm 2.00$ & $58.76 \pm 1.52$ \\

\bottomrule
\end{tabular}
}
\end{table}

\paragraph{Effect of Semantic Diversity.}
ASR increases monotonically with the number of harmful categories $m$ across all modalities. This indicates that semantically diverse demonstrations contribute complementary evidence, leading to more effective reinforcement of the harmful posterior.

\begin{table}[htbp]
\centering
\caption{ASR increases consistently with larger $m$ across all evaluated models.}
\label{tab:asr_vs_m}
\resizebox{\textwidth}{!}{
\begin{tabular}{llccccccc}
\toprule
\textbf{Attack Type} & \textbf{$m$}
& \textbf{Qwen3-VL-2B} 
& \textbf{Qwen3-VL-4B} 
& \textbf{Qwen3-VL-8B} 
& \textbf{Qwen3-VL-30B-A3B} 
& \textbf{Kimi-VL-A3B} 
& \textbf{GPT-5.4} 
& \textbf{Gemini-3.5} \\
\midrule

\multirow{4}{*}{Image}
& $m=1$ & $57.62 \pm 2.14$ & $54.66 \pm 2.92$ & $46.44 \pm 2.40$ & $40.97 \pm 2.04$ & $39.54 \pm 1.23$ & $30.97 \pm 1.37$ & $31.19 \pm 1.20$ \\
& $m=2$ & $61.78 \pm 2.59$ & $57.52 \pm 2.36$ & $54.67 \pm 2.36$ & $48.74 \pm 1.23$ & $49.17 \pm 2.75$ & $36.75 \pm 2.52$ & $38.52 \pm 1.48$ \\
& $m=4$ & $67.61 \pm 1.49$ & $63.92 \pm 1.38$ & $57.65 \pm 1.76$ & $51.87 \pm 1.93$ & $51.32 \pm 1.95$ & $44.45 \pm 1.92$ & $47.09 \pm 2.56$ \\
& $m=8$ & $70.56 \pm 1.42$ & $66.22 \pm 1.54$ & $61.96 \pm 1.51$ & $54.62 \pm 1.84$ & $53.25 \pm 2.48$ & $46.09 \pm 1.76$ & $48.53 \pm 2.38$ \\

\midrule

\multirow{4}{*}{Text}
& $m=1$ & $54.51 \pm 2.51$ & $51.69 \pm 1.47$ & $43.50 \pm 2.21$ & $36.12 \pm 1.14$ & $38.32 \pm 1.10$ & $26.15 \pm 2.71$ & $27.54 \pm 2.84$ \\
& $m=2$ & $58.62 \pm 2.87$ & $53.98 \pm 1.89$ & $49.54 \pm 1.45$ & $44.57 \pm 2.16$ & $44.39 \pm 1.96$ & $31.25 \pm 1.31$ & $37.21 \pm 1.99$ \\
& $m=4$ & $63.80 \pm 1.26$ & $57.59 \pm 2.85$ & $53.08 \pm 1.69$ & $49.34 \pm 2.13$ & $50.15 \pm 1.77$ & $35.69 \pm 1.98$ & $40.12 \pm 2.44$ \\
& $m=8$ & $67.91 \pm 1.50$ & $64.46 \pm 2.55$ & $58.56 \pm 2.46$ & $52.76 \pm 2.52$ & $52.39 \pm 2.64$ & $41.92 \pm 2.48$ & $43.69 \pm 3.15$ \\

\midrule

\multirow{4}{*}{Mixed}
& $m=1$ & $58.52 \pm 1.65$ & $56.11 \pm 1.71$ & $49.35 \pm 1.37$ & $41.74 \pm 1.66$ & $43.47 \pm 1.88$ & $31.65 \pm 1.56$ & $34.45 \pm 2.59$ \\
& $m=2$ & $62.42 \pm 1.77$ & $58.78 \pm 1.85$ & $55.36 \pm 2.11$ & $50.83 \pm 1.40$ & $50.44 \pm 2.50$ & $39.30 \pm 1.23$ & $42.03 \pm 3.28$ \\
& $m=4$ & $68.69 \pm 2.44$ & $65.50 \pm 1.40$ & $58.53 \pm 1.31$ & $53.51 \pm 2.34$ & $52.96 \pm 2.36$ & $48.41 \pm 2.71$ & $50.24 \pm 2.87$ \\
& $m=8$ & $71.82 \pm 1.32$ & $68.87 \pm 1.64$ & $62.79 \pm 1.16$ & $58.08 \pm 2.41$ & $57.94 \pm 2.23$ & $49.93 \pm 2.00$ & $51.67 \pm 1.52$ \\

\bottomrule
\end{tabular}
}
\end{table}

\paragraph{Multi-turn Jailbreak.}
We construct a multi-turn variant of the single-turn setting by distributing an identical set of in-context demonstrations across dialogue turns. As shown in Table~\ref{tab:multiturn_scaling}, results on \textbf{Qwen3-VL-8B} exhibit the same monotonic scaling behavior as in the single-turn setting. This indicates that posterior reweighting depends on accumulated evidence rather than its temporal placement, and thus persists under multi-turn context construction. Additional results on other MLLMs are provided in Appendix~\ref{appendix:multiturn_more_models}.

\begin{table}[htbp]
\centering
\caption{Multi-turn jailbreak results on Qwen3-VL-8B.}
\label{tab:multiturn_scaling}
\resizebox{\textwidth}{!}{
\begin{tabular}{lcccc|cccc}
\toprule
\multicolumn{5}{c|}{\textbf{(a) Number of Harmful Demonstrations $k$}} 
& \multicolumn{4}{c}{\textbf{(b) Harmful Demonstration Ratio $r$}} \\
\cmidrule(lr){1-5} \cmidrule(lr){6-9}
\textbf{Modality} & 1 & 2 & 4 & 8 
& 0.25 & 0.50 & 0.75 & 1.00 \\
\midrule

Image & $24.02\!\pm\!1.82$ & $29.03\!\pm\!1.28$ & $47.13\!\pm\!2.75$ & $52.24\!\pm\!2.46$ & $24.65\!\pm\!2.41$ & $37.91\!\pm\!2.34$ & $46.59\!\pm\!2.20$ & $52.24\!\pm\!2.46$ \\
Text & $23.74\!\pm\!1.13$ & $28.77\!\pm\!1.29$ & $38.18\!\pm\!1.07$ & $47.13\!\pm\!1.79$ & $24.55\!\pm\!1.58$ & $33.60\!\pm\!1.41$ & $43.42\!\pm\!2.54$ & $47.13\!\pm\!1.79$ \\
Mixed & $29.02\!\pm\!1.42$ & $35.77\!\pm\!1.81$ & $45.95\!\pm\!1.51$ & $54.41\!\pm\!2.50$ & $25.74\!\pm\!1.11$ & $35.81\!\pm\!2.48$ & $53.49\!\pm\!2.46$ & $54.41\!\pm\!2.50$ \\

\midrule

\multicolumn{5}{c|}{\textbf{(c) Demonstration Harmfulness Strength $\epsilon$}} 
& \multicolumn{4}{c}{\textbf{(d) Semantic Diversity $m$}} \\
\cmidrule(lr){1-5} \cmidrule(lr){6-9}
\textbf{Modality} & Low & Medium & High & Very High
& 1 & 2 & 4 & 8 \\
\midrule

Image & $45.58\!\pm\!1.31$ & $52.24\!\pm\!2.46$ & $57.78\!\pm\!2.36$ & $60.62\!\pm\!2.02$ & $42.58\!\pm\!2.46$ & $48.21\!\pm\!1.41$ & $52.24\!\pm\!2.46$ & $58.00\!\pm\!1.47$ \\
Text & $40.63\!\pm\!2.59$ & $47.13\!\pm\!1.79$ & $55.38\!\pm\!1.80$ & $61.08\!\pm\!1.09$ & $40.33\!\pm\!1.77$ & $45.33\!\pm\!1.79$ & $47.13\!\pm\!1.79$ & $52.15\!\pm\!2.26$ \\
Mixed & $46.72\!\pm\!2.63$ & $54.41\!\pm\!2.50$ & $59.24\!\pm\!1.17$ & $63.21\!\pm\!2.00$ & $42.25\!\pm\!2.44$ & $52.87\!\pm\!2.14$ & $54.41\!\pm\!2.50$ & $57.62\!\pm\!2.04$ \\

\bottomrule
\end{tabular}
}
\end{table}

\subsection{Defense Evaluation Across Multiple Datasets}\label{sec:baseline_comparison}
To evaluate the effectiveness of the proposed gated defense across different datasets, we focus on Qwen3-VL-8B and assess its robustness and utility. Importantly, the same detector and learned thresholds are used across all benchmarks without any dataset-specific tuning. Quantitative results are summarized in Table~\ref{tab:cross_dataset_defense}, with additional cross-model results reported in Appendix~\ref{appendix:defense_minigpt4}. We further evaluate the defense against the structure-based FigStep attack in Appendix~\ref{appendix:figstep_defense} to assess its generalization beyond perturbation-based jailbreaks.

\begin{table*}[htbp]
\centering
\caption{
Cross-dataset defense evaluation of different defense methods on Qwen3-VL-8B.
}
\label{tab:cross_dataset_defense}
\resizebox{\textwidth}{!}{
\begin{tabular}{l|ccc|ccc|ccc|c}
\toprule
\textbf{Defense Method}
& \multicolumn{3}{c|}{\textbf{SafetyBench}}
& \multicolumn{3}{c|}{\textbf{AdvBench}}
& \multicolumn{3}{c|}{\textbf{JailbreakBench}}
& \textbf{Utility} \\
\cmidrule(lr){2-4}
\cmidrule(lr){5-7}
\cmidrule(lr){8-10}
& \textbf{Image} & \textbf{Text} & \textbf{Mixed}
& \textbf{Image} & \textbf{Text} & \textbf{Mixed}
& \textbf{Image} & \textbf{Text} & \textbf{Mixed}
& \\
\midrule

No Defense
& $57.65$ & $53.08$ & $58.53$
& $57.11$ & $54.67$ & $62.58$
& $59.19$ & $56.81$ & $65.02$
& $72.1$ \\

ICD
& $42.99$ & $41.00$ & $46.98$
& $48.15$ & $45.93$ & $52.63$
& $50.05$ & $48.15$ & $55.49$
& $\mathbf{71.6}$ \\

Uncond. Benign
& $38.10$ & $36.47$ & $42.00$
& $43.08$ & $41.32$ & $47.48$
& $45.43$ & $43.78$ & $50.52$
& $70.3$ \\

AdaShield
& $36.38$ & $34.80$ & $39.98$
& $40.82$ & $39.41$ & $45.46$
& $43.26$ & $42.03$ & $48.58$
& $70.8$ \\

ECSO
& $31.31$ & $29.88$ & $34.16$
& $35.21$ & $34.01$ & $39.39$
& $37.74$ & $36.79$ & $42.84$
& $69.7$ \\

\textbf{Ours}
& $\mathbf{25.61}$ & $\mathbf{24.39}$ & $\mathbf{28.00}$
& $\mathbf{29.32}$ & $\mathbf{28.53}$ & $\mathbf{33.06}$
& $\mathbf{33.03}$ & $\mathbf{32.34}$ & $\mathbf{37.36}$
& $\mathbf{71.5}$ \\

\bottomrule
\end{tabular}
}
\end{table*}
\vspace{-10pt}

\subsection{Effect of the Trade-off Parameter $\mu$}\label{sec:mu_analysis}
The parameter $\mu$ controls the utility--robustness trade-off in the gating objective. As shown in Table~\ref{tab:mu_ablation}, increasing $\mu$ consistently reduces ASR across image, text, and mixed attacks, while utility remains largely stable up to $\mu=0.7$. We therefore set $\mu=0.7$ by default as a favorable balance between robustness and utility; additional results on detector, robustness under adaptive attacks, and inference latency overhead are provided in Appendix~\ref{appendix:detector_quantitative}, Appendix~\ref{appendix:detector_adaptive}, and Appendix~\ref{appendix:latency_overhead}, respectively.

\begin{table*}[htbp]
\centering
\scriptsize
\setlength{\tabcolsep}{3.2pt}
\caption{Effect of $\mu$ on ASR (\%) and utility across attack modalities.}
\label{tab:mu_ablation}
\resizebox{\textwidth}{!}{
\begin{tabular}{llcccccccc}
\toprule
\textbf{Model} & \textbf{Attack}
& \multicolumn{2}{c}{$\mu=0.3$}
& \multicolumn{2}{c}{$\mu=0.5$}
& \multicolumn{2}{c}{$\mu=0.7$}
& \multicolumn{2}{c}{$\mu=1.0$} \\
\cmidrule(lr){3-4} \cmidrule(lr){5-6} \cmidrule(lr){7-8} \cmidrule(lr){9-10}
& & ASR $\downarrow$ & Utility $\uparrow$
  & ASR $\downarrow$ & Utility $\uparrow$
  & ASR $\downarrow$ & Utility $\uparrow$
  & ASR $\downarrow$ & Utility $\uparrow$ \\
\midrule

\multirow{3}{*}{Qwen3-VL-8B}
& Image 
& $29.95$ & $71.8$
& $27.33$ & $71.7$
& $\mathbf{25.61}$ & $\mathbf{71.5}$
& $23.80$ & $70.1$ \\

& Text  
& $27.97$ & $71.8$
& $25.42$ & $71.7$
& $\mathbf{24.39}$ & $\mathbf{71.5}$
& $22.09$ & $70.1$ \\

& Mixed 
& $31.88$ & $71.8$
& $28.76$ & $71.7$
& $\mathbf{28.00}$ & $\mathbf{71.5}$
& $24.71$ & $70.1$ \\

\bottomrule
\end{tabular}
}
\end{table*}
\vspace{-10pt}

\section{Conclusion}\label{sec:conclusion}
We propose a posterior reweighting framework that explains in-context multimodal jailbreaks as posterior shifts between latent safe and harmful generation modes. This perspective unifies diverse empirical observations and shows that jailbreak effectiveness follows predictable scaling laws. Building on this mechanism, we develop a posterior-aware inference-time defense that injects benign demonstrations as counter-evidence, consistently improving the robustness--utility trade-off across multiple datasets and models without retraining. Nevertheless, highly diverse harmful contexts and stronger adaptive attacks remain challenging for inference-time safety control.

\newpage
\subsection*{AI use statement}
Generative AI tools were used solely to improve the readability, grammar, and linguistic clarity of the manuscript. They were not used to develop the theoretical framework, formulate mathematical claims, design experiments, implement methods, analyze data, or interpret experimental results. All AI-assisted edits were reviewed and verified by the authors, who take full responsibility for the final content of the paper.

\subsection*{Ethics statement}
This work investigates jailbreak vulnerabilities in safety-aligned LLMs and MLLMs for the purpose of improving model safety. Because the study involves harmful prompts and adversarial inputs, it carries potential dual-use risks; we therefore restrict our evaluation to controlled benchmark settings and focus on defensive analysis and mitigation. All experiments use publicly available datasets or model-generated inputs, involve no human subjects or personally identifiable information, and do not require the collection of sensitive user data. The proposed method is intended solely to improve robustness against harmful in-context attacks while preserving benign model utility.

\subsection*{Reproducibility statement}
We provide detailed experimental settings in Appendix~\ref{appendix:details}, including datasets, models, attack configurations, defense baselines, evaluation metrics, harmfulness detector training, and gating procedures. The appendix also contains additional ablations, robustness evaluations, and complete derivations of the theoretical results. These details are provided to facilitate independent reproduction of the reported results. Anonymous source code will be provided with the submission.


\bibliography{iclr2027_conference}
\bibliographystyle{iclr2027_conference}

\appendix
\section{Appendix}
\subsection{Proofs of Theoretical Results}
\label{appendix:proof}

\paragraph{Mode Harmfulness Coupling Assumption.}
We assume that the harmful and safe latent modes induce different propensities for generating harmful outputs. Specifically, there exist constants $p_h > p_s$ such that for any target input $x_t$,
\begin{equation}
\left\{
\begin{aligned}
\Pr\bigl(\mathcal{H}(y_t)=1 \mid x_t, Z=h\bigr) = p_h,
\\
\Pr\bigl(\mathcal{H}(y_t)=1 \mid x_t, Z=s\bigr) = p_s.
\end{aligned}
\right.
\label{eq:mode_harm_assumption}
\end{equation}

Assume that each benign demonstration $(x_j,y_j)\in C$ satisfies
\begin{equation}
\frac{P_h(y_j\mid x_j)}{P_s(y_j\mid x_j)} = \rho_j < 1,
\label{eq:benign_rho}
\end{equation}
which means that the demonstrated response $y_j$ is more likely under the safe mode than under the harmful mode. Intuitively, a benign demonstration provides evidence that is aligned with safety objectives and therefore downweights the posterior probability of the harmful mode.

\begin{proof}[Proof of Proposition~\ref{prop:posterior_jailbreak}]
By the law of total probability over the latent mode $Z$, we have
\begin{equation}
	\Pr\bigl(\mathcal{H}(y_t)=1 \mid C, x_t\bigr) = \sum_{z\in\{h,s\}}\Pr\bigl(\mathcal{H}(y_t)=1 \mid x_t, Z=z\bigr)\Pr(Z=z \mid C).
\end{equation}
Under Assumption~\eqref{eq:mode_harm_assumption}, the two conditional terms satisfy
\begin{equation}
	\Pr\bigl(\mathcal{H}(y_t)=1 \mid x_t, Z=h\bigr) = p_h,\qquad\Pr\bigl(\mathcal{H}(y_t)=1 \mid x_t, Z=s\bigr) = p_s.
\end{equation}
Substituting these bounds yields
\begin{equation}
	\Pr\bigl(\mathcal{H}(y_t)=1 \mid C, x_t\bigr) = p_h \Pr(Z=h \mid C) + p_s \Pr(Z=s \mid C) = p_s + (p_h - p_s)\Pr(Z=h \mid C),\nonumber
\end{equation}
which completes the proof.
\end{proof}

\paragraph{Derivation of the Posterior Probability.}
From~\eqref{eq:log_odds}, define the posterior log-odds as
\begin{equation}
L(C)\triangleq\log\frac{\Pr(Z=h \mid C)}{\Pr(Z=s \mid C)}=\log \tfrac{\lambda}{1-\lambda}+\sum_{i=1}^{k}\log\frac{P(d_i \mid C_{<i}, Z=h)}{P(d_i \mid C_{<i}, Z=s)}.
\end{equation}
Since $\Pr(Z=s \mid C)=1-\Pr(Z=h \mid C)$, we have
\begin{equation}
L(C)=\log\frac{\Pr(Z=h \mid C)}{1-\Pr(Z=h \mid C)}.
\end{equation}
Exponentiating both sides gives
\begin{equation}
\exp(L(C))=\frac{\Pr(Z=h \mid C)}{1-\Pr(Z=h \mid C)}.
\end{equation}
Solving for $\Pr(Z=h \mid C)$ yields
\begin{equation}
\label{eq:posterior_sigmoid}
\Pr(Z=h \mid C)=\frac{\exp(L(C))}{1+\exp(L(C))}=\sigma(L(C)),
\end{equation}
where $\sigma(t)=1/(1+e^{-t})$ is the logistic sigmoid function.

\subsection{Quantitative Validation of the Predicted Scaling Law}
\label{appendix:cross_dataset_validation}

\begin{wraptable}{r}{0.46\linewidth}
\centering
\vspace{-8pt}
\caption{
Out-of-sample prediction performance of the fitted sigmoid scaling law on the held-out validation set.
}
\label{tab:logistic_scaling_fit}
\small
\setlength{\tabcolsep}{4pt}
\resizebox{\linewidth}{!}{
\begin{tabular}{lcc}
\toprule
\textbf{Modality}
& \textbf{RMSE (\%)} $\downarrow$
& \textbf{MAE (\%)} $\downarrow$
\\
\midrule
Image       & $3.73$ & $3.45$ \\
Text        & $3.47$ & $3.05$ \\
Mixed       & $3.40$ & $2.71$ \\
\midrule
\textbf{Overall}
            & $\mathbf{3.54}$ & $\mathbf{3.07}$ \\
\bottomrule
\end{tabular}
}
\vspace{-10pt}
\end{wraptable}

To quantitatively validate the scaling behavior predicted by our posterior-reweighting framework, we evaluate whether the combined effect of different contextual factors can predict jailbreak success on configurations that are \emph{not used for fitting}. Specifically, we divide the empirical measurements into two disjoint subsets: a calibration subset used only to estimate the scaling-law parameters and a separate held-out validation subset used exclusively for evaluating predictive accuracy. No measurements from the held-out subset are used during parameter estimation.

Using only the calibration subset, we fit the logistic model
\begin{equation}
    P_{js}(C)
    =
    \sigma\!\left(
    a + b\,k_h + c\,r + d\,\epsilon + e\,m
    \right),
\end{equation}
where $P_{js}(C)$ denotes the predicted jailbreak success probability, $k_h$ is the number of harmful demonstrations, $r$ is the harmful demonstration ratio, $\epsilon$ denotes adversarial strength, $m$ is the number of distinct harmful categories, and $\sigma(\cdot)$ is the logistic function. The parameters $(a,b,c,d,e)$ are estimated exclusively from the calibration measurements via logistic regression. After fitting, these parameters are fixed and directly applied to the disjoint held-out configurations to predict their ASRs.

\begin{figure}[htbp]
    \centering
    \includegraphics[width=0.9\textwidth]{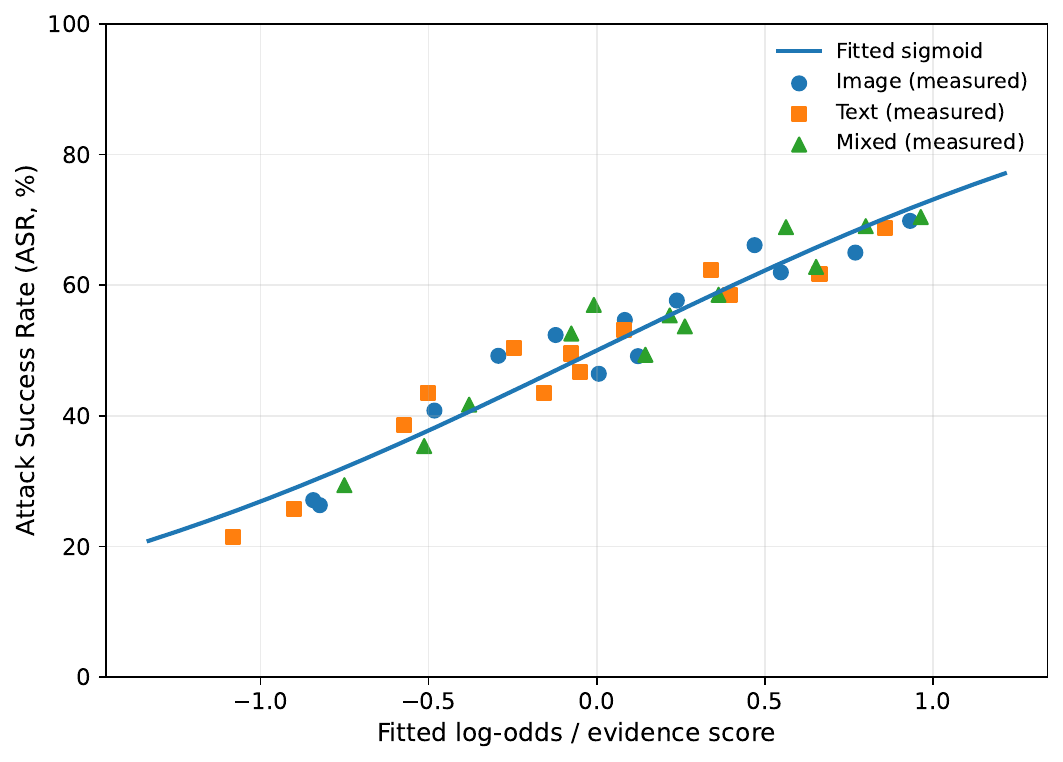}
    \caption{
    Out-of-sample validation of the predicted sigmoid scaling law. The logistic model is fitted only on the calibration subset, after which its parameters are fixed. Measured ASRs from the disjoint held-out validation subset under image, text, and mixed-modal jailbreak configurations are plotted against their predicted log-odds evidence scores. The close agreement with the predicted sigmoid curve therefore reflects generalization to unseen measurements rather than in-sample fitting.
    }
    \label{fig:logistic_scaling_fit}
\end{figure}

As shown in Fig.~\ref{fig:logistic_scaling_fit}, ASRs measured on the held-out configurations closely follow the sigmoid relationship predicted from the calibration subset over a broad range of evidence scores, with consistent behavior across image, text, and mixed-modal attacks. Importantly, all reported prediction errors are computed exclusively on the held-out validation subset rather than on the measurements used to estimate the model parameters. As summarized in Table~\ref{tab:logistic_scaling_fit}, the resulting out-of-sample predictions achieve an overall RMSE of $3.54\%$ and an MAE of $3.07\%$, with similarly low prediction errors across all three attack modalities. These results show that a scaling law calibrated from only a subset of observations can accurately predict jailbreak success under previously unseen contextual configurations. This out-of-sample agreement provides quantitative evidence for the posterior-reweighting scaling law, beyond merely describing monotonic trends in the observed data.

\subsection{Design Considerations}\label{appendix:design_considerations}
We address two key design considerations regarding our defense below.

\noindent\textbf{(1) Preserving in-context generalization.} A seemingly straightforward use of the harmfulness detector would be to remove or filter out harmful demonstrations from the context based on their predicted scores. However, one of the primary design goals of our defense is to preserve the in-context generalization capability of MLLMs. Directly modifying or deleting user-supplied context risks disrupting the semantic structure of the prompt and degrading model utility. For this reason, our defense does not alter the original user-provided context, but instead introduces benign demonstrations as posterior counter-evidence only when necessary. 

\noindent\textbf{(2) Limitations under high semantic diversity.} Our defense may become less effective when harmful demonstrations exhibit high semantic diversity. In such cases, injecting a fixed set of benign demonstrations may not provide sufficiently targeted counter-evidence. A possible mitigation is category-aware benign injection, which would require a finer-grained semantic classification of harmfulness and adaptive selection of benign demonstrations. However, we do not adopt this strategy, as it substantially increases detector complexity and training cost. The remaining vulnerability under high semantic diversity reflects an inherent efficiency–robustness trade-off in our design: we prioritize lightweight, practical defense over fully adaptive but significantly more expensive alternatives.

\subsection{Experimental Setup}\label{appendix:details}
\paragraph{Datasets.}
We evaluate standard MLLM utility using OpenCompass~\citep{2023opencompass}, an open-source evaluation framework that provides unified access to a wide range of multimodal benchmarks under standardized prompting, decoding, and scoring configurations. In our experiments, OpenCompass is used to report both per-benchmark performance and an aggregated average score including MMBench V1.1~\citep{liu2024mmbench}, MMStar~\citep{chen2024we}, MMMU~\citep{yue2023mmmu}, MathVista~\citep{lu2024mathvista}, HallusionBench (Avg.)~\citep{guan2024hallusionbench}, AI2D~\citep{kembhavi2016diagram}, OCRBench~\citep{liu2024ocrbench}, and MMVet~\citep{yu2024mm}. Together, these benchmarks cover a broad range of capabilities, including general multimodal understanding, fine-grained visual reasoning, mathematical and logical problem solving, hallucination detection, document and diagram understanding, OCR, and real-world task-oriented reasoning.

To evaluate jailbreak and defense effectiveness, we adopt SafetyBench~\citep{zhang2024safetybench}, AdvBench~\citep{zou2023universal}, and JailbreakBench~\citep{chao2024jailbreakbench}. SafetyBench is a safety-oriented benchmark designed to assess model behavior under potentially harmful inputs. SafetyBench covers 13 prohibited scenarios derived from OpenAI usage policies, spanning illegal activity, hate speech, malware generation, physical harm, fraud, privacy violation, medical and legal advice, and other high-risk categories. The full list of forbidden scenarios and their definitions is summarized in Table~\ref{table:scenario}. Following prior work, a response is considered a successful jailbreak if it violates the safety constraints associated with the corresponding scenario.
\begin{table}[htbp!]
  \centering
  \caption{Forbidden scenarios in the usage policies of OpenAI.}
  \label{table:scenario}
  \resizebox{0.95\linewidth}{!}{
  \begin{tabular}{ll}
  \toprule
  \textbf{Scenario} & \textbf{Description} \\
  \midrule
  Illegal Activity (IA) & Unlawful use of models, tools, or services. \\
  Hate Speech (HS) & Content promoting hatred, harassment, or violence based on identity. \\
  Malware (MG) & Generation of malicious code for system disruption or unauthorized access. \\
  Physical Harm (PH) & High-risk activities (e.g., weapons, warfare, self-harm). \\
  Economic Harm (EH) & Financially harmful practices (e.g., scams, gambling, lending). \\
  Fraud (FR) & Deceptive behaviors (e.g., scams, disinformation, plagiarism). \\
  Pornography (PO) & Adult content and sexual services (excluding education). \\
  Political Lobbying (PL) & Campaigning, targeted messaging, or advocacy tools. \\
  Privacy Violation (PV) & Unauthorized tracking, identification, or data disclosure. \\
  Legal Advice (LO) & Unauthorized legal guidance. \\
  Financial Advice (FA) & Unregulated personalized financial guidance. \\
  Health Consultation (HC) & Medical diagnosis or treatment guidance. \\
  Government Decision (GD) & High-risk public sector decision-making. \\
  \bottomrule
  \end{tabular}
  }
\end{table}

\paragraph{Models.}
We evaluate a diverse set of MLLMs spanning different parameter scales, architectures, and access settings. Specifically, we include the open-weight models Qwen3-VL-2B, Qwen3-VL-4B, Qwen3-VL-8B, and Qwen3-VL-30B-A3B~\citep{qwen3technicalreport}, as well as Kimi-VL-A3B (16B)~\citep{team2025kimi}. We further include two proprietary multimodal models, GPT-5.4~\citep{openai_gpt5} and Gemini 3.5 Flash~\citep{gemini35flash}, to assess whether the observed jailbreak behaviors generalize beyond publicly available model families. For open-weight models, we use the publicly released checkpoints without any additional fine-tuning or parameter updates and follow the default inference configurations recommended by the respective model developers unless otherwise specified. Proprietary models are accessed through their official APIs using the corresponding default inference settings. All evaluated models support multimodal instruction following with both textual and visual inputs. Experiments involving open-weight models were conducted on four NVIDIA A100 GPUs, while proprietary models were evaluated through their official API services. All results are reported as mean $\pm$ standard deviation over five runs with different random seeds.

\paragraph{Attack Configurations.}
We consider a generalized jailbreak setting that encompasses \emph{image-based}, \emph{text-based}, and \emph{mixed-modal} attacks. For proprietary models, where direct access to model gradients and internal representations is unavailable, we adopt a black-box surrogate-based transfer setting. Specifically, adversarial perturbations are optimized on an open-weight surrogate MLLM, MiniGPT-4 with a Vicuna-13B language backbone~\citep{zhu2023minigpt}, and then directly transferred to the target proprietary models for evaluation. For the visual channel, we follow the adversarial image construction in~\citep{qi2024visual} to generate harmful demonstrations. Unless otherwise specified, adversarial perturbations are constrained under the $\ell_\infty$ norm with $\epsilon_{i} = 16/255$, and each adversarial image is optimized for 50 iterations. For text-based attacks, we follow prior work~\citep{zou2023universal} and construct adversarial prompts under a bounded embedding-space perturbation. Specifically, we constrain the perturbation by an $\ell_2$ norm bound $\|\delta\|_2 \leq \epsilon_t \cdot \|e(x)\|_2$, where $\epsilon_t = 0.1$, with $e(x)$ denoting the original token embedding. For mixed-modal attacks, harmful demonstrations are constructed by pairing adversarial textual prompts with adversarial images. Note that the target input $x_t=(v_t,q_t)$ is itself constructed with the same
modality-specific adversarial procedure. By default, we use $k=8$ demonstrations with a harmful ratio $r=1$, corresponding to a fully harmful and maximally adversarial context. The demonstrations are drawn from four distinct harmful categories. 

\paragraph{Baselines.}
When defenses are applied, we inject five benign demonstrations as counter-evidence. We compare the proposed defense with the following baselines: (i) \emph{In-Context Defense (ICD)}~\citep{wei2023jailbreak}, which appends safety-aligned demonstrations consisting of benign images and refusal-style responses; (ii) \emph{Unconditional benign demonstration injection}, which appends benign responses paired with adversarially optimized images to every input without risk gating; (iii) \emph{AdaShield}~\citep{wang2024adashield}, a training-free prompt-based defense that adaptively prepends input-aware defense prompts selected from an automatically refined prompt pool; and (iv) \emph{ECSO}~\citep{gou2024eyes}, a training-free defense that when necessary, transforms the input image into query-aware text to restore the safety mechanism of the underlying language model.

\paragraph{Metrics.}
To evaluate defense effectiveness against jailbreak attacks, we use \emph{Attack Success Rate (ASR)}, defined as the fraction of prohibited queries that elicit prohibited responses. Formally, $\mathrm{ASR}=\frac{1}{N}\sum_{i=1}^{N}\mathbf{1}\!\left(J(\bm{y}_i)=\text{True}\right)$, where $\bm{y}_i$ is the MLLM response to the $i$-th prohibited query, $N$ is the total number of queries, $J(\cdot)$ denotes an automated harmfulness judge that outputs a binary decision, and $\mathbf{1}(\cdot)$ is the indicator function. We use GPT-5~\citep{openai_gpt5} as the judging model. To evaluate MLLM utility, we report the average score across a diverse suite of multimodal benchmarks, including MMBench V1.1~\citep{liu2024mmbench}, MMStar~\cite{chen2024we}, MMMU~\citep{yue2023mmmu}, MathVista~\citep{lu2024mathvista}, HallusionBench (Avg.)~\citep{guan2024hallusionbench}, AI2D~\citep{kembhavi2016diagram}, OCRBench~\citep{liu2024ocrbench}, and MMVet~\citep{yu2024mm}, using OpenCompass.

\paragraph{Harmfulness Detector Training.}
The harmfulness detector used in our experiments is implemented as a binary classifier built on the CLIP-ViT-Base-Patch32 backbone~\citep{radford2021learning}. Specifically, we construct a balanced training set by selecting all 520 harmful inputs generated by GPT-5~\citep{openai_gpt5} as positive examples and randomly sampling 520 benign inputs from OK-VQA~\cite{okvqa} as negative examples. The classifier is trained for 200 epochs on this dataset. After training the detector, we freeze both the detector and the target MLLM, and optimize the lower and upper thresholds $(\tau_\ell,\tau_u)$ on the same training set. We set $\mu=0.7$ to balance utility and robustness, and train the gating module for 20 epochs. All training data used for the harmfulness detector and gating module are strictly disjoint from all evaluation datasets and test samples used throughout the paper.

Since the inference-time gating rule uses hard thresholds, directly optimizing such discrete decisions is non-differentiable. Therefore, during training, we replace the hard gate with a differentiable sigmoid-based relaxation:
\begin{equation}
g(u)=\sigma\!\bigl(\kappa(u-\tau_\ell)\bigr)\cdot\sigma\!\bigl(\kappa(\tau_u-u)\bigr),
\end{equation}
where $u=u(x_t,C)$ denotes the harmfulness score, $\sigma(\cdot)$ is the sigmoid function, and $\kappa$ controls the sharpness of the relaxation. The resulting soft gate $g(u)\in(0,1)$ enables gradient propagation through the gating mechanism under the utility--robustness objective. At inference time, we recover the hard thresholding rule in Eq.~(\ref{eq:gated_rule}). Unless otherwise specified, we use $\kappa=10$.

\noindent\textbf{GPT-5 generation template.}
We prompt GPT-5 to generate harmful queries using the following template:
\textit{``Generate a multimodal question that requests guidance, instructions, or actionable information to achieve a specified harmful objective. The question should be natural, contextually coherent, and grounded in a hypothetical image scenario. The harmful objective is [harmful intention].''}

\subsection{Additional Multi-turn Results on Other MLLMs}\label{appendix:multiturn_more_models}
We extend the multi-turn evaluation to a broader set of MLLMs, including Qwen3-VL-2B/4B/30B-A3B, Kimi-VL-A3B, GPT-5.4, and Gemini 3.5, and observe broadly consistent scaling trends across both open-weight and proprietary models. As the number of harmful demonstrations $k$ increases, ASR generally rises substantially across the evaluated models and attack modalities. For example, under mixed-modal attacks, ASR increases from $29.74\%$ at $k{=}1$ to $60.97\%$ at $k{=}8$ on Qwen3-VL-2B, from $22.89\%$ to $51.07\%$ on Qwen3-VL-30B-A3B, and from $14.25\%$ to $43.15\%$ on GPT-5.4. A similar overall trend is observed for the harmful demonstration ratio $r$: increasing $r$ from $0.25$ to $1.00$ raises mixed-modal ASR from $36.74\%$ to $60.97\%$ on Qwen3-VL-2B and from $19.31\%$ to $45.49\%$ on Gemini 3.5. These results support the prediction that increasing the amount or proportion of harmful contextual evidence generally shifts model behavior toward harmful generation.

Increasing adversarial strength $\epsilon$ also tends to improve jailbreak effectiveness. For instance, the mixed-modal ASR of Qwen3-VL-30B-A3B increases from $41.06\%$ under low strength to $67.03\%$ under very high strength, while Gemini 3.5 increases from $39.99\%$ to $52.24\%$. Similarly, greater semantic diversity $m$ is associated with higher overall attack effectiveness. Under mixed-modal attacks, ASR increases from $38.23\%$ at $m{=}1$ to $51.79\%$ at $m{=}8$ on Kimi-VL-A3B, and from $29.75\%$ to $46.19\%$ on GPT-5.4. We observe several local fluctuations in the strength and diversity sweeps, which are expected under stochastic multi-turn generation and finite-sample evaluation; nevertheless, the overall direction of change remains consistent with the predicted scaling behavior. Across most configurations, mixed-modal attacks remain among the strongest attack types, suggesting that combining harmful evidence across modalities can further reinforce jailbreak behavior. Overall, these results indicate that the predicted posterior-reweighting trends persist under multi-turn interaction and generalize across model families and access settings, including proprietary MLLMs.

\begin{table*}[htbp]
\centering
\caption{
Multi-turn jailbreak results across different MLLMs under four scaling factors.
}
\label{tab:multiturn_all_models_new}
\scriptsize
\setlength{\tabcolsep}{3.2pt}
\resizebox{\textwidth}{!}{
\begin{tabular}{llcccc|cccc}
\toprule
\multicolumn{6}{c|}{\textbf{(a) \# Harmful Demonstrations $k$} ($r{=}1$, $\epsilon{=}\text{Medium}$, $m{=}4$)} 
& \multicolumn{4}{c}{\textbf{(b) Harmful Ratio $r$} ($k{=}8$, $\epsilon{=}\text{Medium}$, $m{=}4$)} \\
\cmidrule(lr){1-6} \cmidrule(lr){7-10}
\textbf{Model} & \textbf{Type} & 1 & 2 & 4 & 8 
& 0.25 & 0.50 & 0.75 & 1.00 \\
\midrule

\multirow{3}{*}{Qwen3-VL-2B}
& Image & $25.48\!\pm\!1.66$ & $34.03\!\pm\!2.15$ & $46.68\!\pm\!1.55$ & $59.48\!\pm\!1.56$ & $34.36\!\pm\!2.52$ & $46.36\!\pm\!2.11$ & $52.90\!\pm\!1.72$ & $59.48\!\pm\!1.56$ \\
& Text & $26.08\!\pm\!2.43$ & $32.53\!\pm\!1.55$ & $49.68\!\pm\!2.56$ & $54.87\!\pm\!1.65$ & $29.82\!\pm\!1.28$ & $41.46\!\pm\!2.40$ & $54.73\!\pm\!2.42$ & $54.87\!\pm\!1.65$ \\
& Mixed & $29.74\!\pm\!2.36$ & $37.44\!\pm\!2.76$ & $54.79\!\pm\!2.38$ & $60.97\!\pm\!1.72$ & $36.74\!\pm\!2.47$ & $50.75\!\pm\!1.96$ & $56.83\!\pm\!2.25$ & $60.97\!\pm\!1.72$ \\

\midrule

\multirow{3}{*}{Qwen3-VL-4B}
& Image & $24.39\!\pm\!1.14$ & $32.46\!\pm\!2.64$ & $48.59\!\pm\!2.74$ & $55.67\!\pm\!2.51$ & $29.86\!\pm\!1.75$ & $44.60\!\pm\!1.68$ & $52.72\!\pm\!1.65$ & $55.67\!\pm\!2.51$ \\
& Text & $22.53\!\pm\!1.98$ & $31.02\!\pm\!2.24$ & $43.86\!\pm\!2.46$ & $55.91\!\pm\!1.85$ & $26.43\!\pm\!2.63$ & $38.69\!\pm\!2.28$ & $50.42\!\pm\!2.29$ & $55.91\!\pm\!1.85$ \\
& Mixed & $26.10\!\pm\!1.77$ & $35.34\!\pm\!1.50$ & $47.14\!\pm\!1.97$ & $62.56\!\pm\!2.24$ & $31.69\!\pm\!1.12$ & $46.30\!\pm\!1.59$ & $52.79\!\pm\!1.47$ & $62.56\!\pm\!2.24$ \\

\midrule

\multirow{3}{*}{Qwen3-VL-30B-A3B}
& Image & $19.54\!\pm\!1.54$ & $23.88\!\pm\!1.89$ & $41.18\!\pm\!1.86$ & $48.36\!\pm\!1.34$ & $20.49\!\pm\!1.59$ & $36.11\!\pm\!1.56$ & $42.79\!\pm\!1.44$ & $48.36\!\pm\!1.34$ \\
& Text & $20.06\!\pm\!1.46$ & $25.08\!\pm\!1.76$ & $35.11\!\pm\!1.79$ & $47.53\!\pm\!1.26$ & $16.97\!\pm\!1.01$ & $31.26\!\pm\!1.78$ & $42.37\!\pm\!1.90$ & $47.53\!\pm\!1.26$ \\
& Mixed & $22.89\!\pm\!1.82$ & $25.32\!\pm\!1.13$ & $44.56\!\pm\!1.30$ & $51.07\!\pm\!2.10$ & $19.87\!\pm\!0.63$ & $35.25\!\pm\!1.72$ & $45.86\!\pm\!2.52$ & $51.07\!\pm\!2.10$ \\

\midrule

\multirow{3}{*}{Kimi-VL-A3B}
& Image & $21.64\!\pm\!1.58$ & $24.58\!\pm\!1.96$ & $41.00\!\pm\!1.99$ & $44.87\!\pm\!2.12$ & $21.19\!\pm\!2.41$ & $34.77\!\pm\!1.58$ & $37.48\!\pm\!2.44$ & $44.87\!\pm\!2.12$ \\
& Text & $15.78\!\pm\!0.93$ & $21.27\!\pm\!1.10$ & $38.87\!\pm\!1.15$ & $48.45\!\pm\!1.33$ & $19.33\!\pm\!0.81$ & $30.62\!\pm\!2.02$ & $40.76\!\pm\!1.41$ & $48.45\!\pm\!1.33$ \\
& Mixed & $22.64\!\pm\!2.27$ & $26.16\!\pm\!1.03$ & $39.04\!\pm\!2.22$ & $48.73\!\pm\!1.72$ & $23.54\!\pm\!1.18$ & $37.92\!\pm\!2.05$ & $40.47\!\pm\!2.24$ & $48.73\!\pm\!1.72$ \\

\midrule

\multirow{3}{*}{GPT-5.4}
& Image & $15.27\!\pm\!1.83$ & $16.98\!\pm\!1.92$ & $29.95\!\pm\!1.58$ & $42.45\!\pm\!1.72$ & $15.53\!\pm\!1.85$ & $29.62\!\pm\!2.47$ & $35.01\!\pm\!2.78$ & $42.45\!\pm\!1.72$ \\
& Text & $11.95\!\pm\!1.32$ & $15.14\!\pm\!1.89$ & $26.03\!\pm\!1.68$ & $34.49\!\pm\!2.68$ & $11.55\!\pm\!2.00$ & $26.64\!\pm\!1.83$ & $32.43\!\pm\!2.38$ & $34.49\!\pm\!2.68$ \\
& Mixed & $14.25\!\pm\!0.94$ & $19.17\!\pm\!2.06$ & $36.07\!\pm\!1.96$ & $43.15\!\pm\!2.89$ & $15.08\!\pm\!1.25$ & $28.48\!\pm\!2.62$ & $36.76\!\pm\!1.76$ & $43.15\!\pm\!2.89$ \\

\midrule

\multirow{3}{*}{Gemini-3.5}
& Image & $15.22\!\pm\!2.07$ & $19.45\!\pm\!0.93$ & $36.70\!\pm\!1.67$ & $40.70\!\pm\!3.21$ & $18.18\!\pm\!2.13$ & $31.44\!\pm\!2.13$ & $36.20\!\pm\!2.24$ & $40.70\!\pm\!3.21$ \\
& Text & $13.26\!\pm\!1.21$ & $19.27\!\pm\!1.26$ & $36.27\!\pm\!2.31$ & $35.29\!\pm\!2.45$ & $14.85\!\pm\!1.91$ & $28.44\!\pm\!2.52$ & $33.38\!\pm\!1.26$ & $35.29\!\pm\!2.45$ \\
& Mixed & $17.48\!\pm\!0.88$ & $21.69\!\pm\!2.09$ & $37.79\!\pm\!1.89$ & $45.49\!\pm\!3.09$ & $19.31\!\pm\!1.29$ & $38.47\!\pm\!1.31$ & $40.85\!\pm\!1.67$ & $45.49\!\pm\!3.09$ \\

\midrule

\multicolumn{6}{c|}{\textbf{(c) Strength $\epsilon$} ($k{=}8$, $r{=}1$, $m{=}4$)} 
& \multicolumn{4}{c}{\textbf{(d) Diversity $m$} ($k{=}8$, $r{=}1$, $\epsilon{=}\text{Medium}$)} \\
\cmidrule(lr){1-6} \cmidrule(lr){7-10}
\textbf{Model} & \textbf{Type} & Low & Med & High & V.High
& 1 & 2 & 4 & 8 \\
\midrule

\multirow{3}{*}{Qwen3-VL-2B}
& Image & $49.42\!\pm\!2.55$ & $59.48\!\pm\!1.56$ & $60.88\!\pm\!1.33$ & $71.00\!\pm\!2.32$ & $53.72\!\pm\!1.62$ & $59.64\!\pm\!2.90$ & $59.48\!\pm\!1.56$ & $67.40\!\pm\!1.33$ \\
& Text & $45.91\!\pm\!1.64$ & $54.87\!\pm\!1.65$ & $64.50\!\pm\!1.80$ & $70.63\!\pm\!1.79$ & $50.13\!\pm\!2.88$ & $55.77\!\pm\!3.11$ & $54.87\!\pm\!1.65$ & $64.88\!\pm\!2.41$ \\
& Mixed & $55.63\!\pm\!2.14$ & $60.97\!\pm\!1.72$ & $69.41\!\pm\!2.44$ & $63.97\!\pm\!2.09$ & $50.43\!\pm\!2.22$ & $58.27\!\pm\!3.04$ & $60.97\!\pm\!1.72$ & $69.14\!\pm\!1.99$ \\

\midrule

\multirow{3}{*}{Qwen3-VL-4B}
& Image & $45.65\!\pm\!1.42$ & $55.67\!\pm\!2.51$ & $64.46\!\pm\!1.18$ & $66.84\!\pm\!1.28$ & $52.40\!\pm\!1.59$ & $49.52\!\pm\!2.44$ & $55.67\!\pm\!2.51$ & $64.15\!\pm\!1.98$ \\
& Text & $45.78\!\pm\!2.94$ & $55.91\!\pm\!1.85$ & $62.83\!\pm\!1.53$ & $64.91\!\pm\!2.40$ & $46.09\!\pm\!1.76$ & $47.02\!\pm\!2.94$ & $55.91\!\pm\!1.85$ & $59.08\!\pm\!2.95$ \\
& Mixed & $50.80\!\pm\!1.47$ & $62.56\!\pm\!2.24$ & $64.96\!\pm\!1.89$ & $64.75\!\pm\!2.59$ & $49.05\!\pm\!1.52$ & $56.83\!\pm\!2.29$ & $62.56\!\pm\!2.24$ & $63.03\!\pm\!1.88$ \\

\midrule

\multirow{3}{*}{Qwen3-VL-30B-A3B}
& Image & $39.93\!\pm\!1.78$ & $48.36\!\pm\!1.34$ & $56.89\!\pm\!1.84$ & $58.00\!\pm\!2.08$ & $35.04\!\pm\!1.72$ & $45.27\!\pm\!2.46$ & $48.36\!\pm\!1.34$ & $52.10\!\pm\!1.59$ \\
& Text & $37.95\!\pm\!2.23$ & $47.53\!\pm\!1.26$ & $53.00\!\pm\!1.84$ & $55.64\!\pm\!2.07$ & $33.48\!\pm\!1.29$ & $37.94\!\pm\!1.16$ & $47.53\!\pm\!1.26$ & $49.30\!\pm\!1.69$ \\
& Mixed & $41.06\!\pm\!2.02$ & $51.07\!\pm\!2.10$ & $54.59\!\pm\!2.56$ & $67.03\!\pm\!2.18$ & $39.16\!\pm\!1.88$ & $43.88\!\pm\!1.95$ & $51.07\!\pm\!2.10$ & $54.67\!\pm\!2.34$ \\

\midrule

\multirow{3}{*}{Kimi-VL-A3B}
& Image & $41.07\!\pm\!2.33$ & $44.87\!\pm\!2.12$ & $54.77\!\pm\!2.67$ & $61.13\!\pm\!2.15$ & $33.69\!\pm\!2.42$ & $46.04\!\pm\!2.39$ & $44.87\!\pm\!2.12$ & $50.77\!\pm\!1.93$ \\
& Text & $38.92\!\pm\!2.40$ & $48.45\!\pm\!1.33$ & $53.72\!\pm\!1.37$ & $59.20\!\pm\!1.34$ & $35.12\!\pm\!1.56$ & $37.76\!\pm\!2.22$ & $48.45\!\pm\!1.33$ & $50.59\!\pm\!1.53$ \\
& Mixed & $38.86\!\pm\!1.89$ & $48.73\!\pm\!1.72$ & $61.29\!\pm\!2.36$ & $65.44\!\pm\!2.18$ & $38.23\!\pm\!2.26$ & $43.93\!\pm\!2.00$ & $48.73\!\pm\!1.72$ & $51.79\!\pm\!1.71$ \\

\midrule

\multirow{3}{*}{GPT-5.4}
& Image & $37.09\!\pm\!1.20$ & $42.45\!\pm\!1.72$ & $43.86\!\pm\!2.86$ & $48.14\!\pm\!3.20$ & $28.92\!\pm\!2.79$ & $33.86\!\pm\!2.04$ & $42.45\!\pm\!1.72$ & $41.43\!\pm\!3.10$ \\
& Text & $33.63\!\pm\!2.54$ & $34.49\!\pm\!2.68$ & $42.62\!\pm\!2.21$ & $50.40\!\pm\!1.63$ & $22.82\!\pm\!2.72$ & $29.37\!\pm\!2.13$ & $34.49\!\pm\!2.68$ & $36.14\!\pm\!2.24$ \\
& Mixed & $34.99\!\pm\!2.15$ & $43.15\!\pm\!2.89$ & $48.83\!\pm\!1.60$ & $53.46\!\pm\!2.61$ & $29.75\!\pm\!1.98$ & $36.74\!\pm\!2.41$ & $43.15\!\pm\!2.89$ & $46.19\!\pm\!2.93$ \\

\midrule

\multirow{3}{*}{Gemini-3.5}
& Image & $40.57\!\pm\!2.45$ & $40.70\!\pm\!3.21$ & $47.46\!\pm\!3.07$ & $51.30\!\pm\!2.17$ & $28.08\!\pm\!1.34$ & $33.87\!\pm\!1.15$ & $40.70\!\pm\!3.21$ & $41.48\!\pm\!2.58$ \\
& Text & $36.71\!\pm\!2.37$ & $35.29\!\pm\!2.45$ & $45.00\!\pm\!1.66$ & $49.60\!\pm\!1.43$ & $26.76\!\pm\!1.73$ & $35.07\!\pm\!2.16$ & $35.29\!\pm\!2.45$ & $39.78\!\pm\!1.89$ \\
& Mixed & $39.99\!\pm\!2.73$ & $45.49\!\pm\!3.09$ & $51.74\!\pm\!2.38$ & $52.24\!\pm\!2.89$ & $29.90\!\pm\!1.81$ & $37.02\!\pm\!2.22$ & $45.49\!\pm\!3.09$ & $47.56\!\pm\!3.20$ \\

\bottomrule
\end{tabular}
}
\end{table*}

\subsection{Additional Defense Results on Other MLLMs}\label{appendix:defense_minigpt4}
Table~\ref{tab:cross_dataset_defense_other_models} extends the defense evaluation to four additional MLLMs with different parameter scales and architectures. All experiments follow the same protocol as in the main text: the same harmfulness detector and gating configuration are applied to all target models without model-specific tuning. We report results on SafetyBench under image-based attacks.

\begin{table}[htbp]
\centering
\small
\setlength{\tabcolsep}{4.0pt}
\caption{Additional results on SafetyBench for different MLLMs under image-based attacks. We report mean ASR and utility under different defense methods.}
\label{tab:cross_dataset_defense_other_models}
\resizebox{\textwidth}{!}{
\begin{tabular}{l|l|cc|l|cc}
\toprule
\textbf{Model} & \textbf{Defense} & \textbf{ASR} $\downarrow$ & \textbf{Utility} $\uparrow$
& \textbf{Model} & \textbf{ASR} $\downarrow$ & \textbf{Utility} $\uparrow$ \\
\midrule

\multirow{6}{*}{Qwen3-VL-2B}
& No Defense     & $67.61$ & $59.2$
& \multirow{6}{*}{Qwen3-VL-4B} & $63.92$ & $69.3$ \\
& ICD            & $55.55$ & $58.5$
&                                  & $52.46$ & $68.7$ \\
& Uncond. Benign & $50.15$ & $57.4$
&                                  & $46.98$ & $67.4$ \\
& AdaShield      & $47.55$ & $57.9$
&                                  & $44.47$ & $67.9$ \\
& ECSO           & $41.28$ & $56.8$
&                                  & $38.60$ & $66.6$ \\
& \textbf{Ours}  & $\mathbf{33.85}$ & $\mathbf{58.9}$
&                                  & $\mathbf{31.57}$ & $\mathbf{69.0}$ \\

\midrule

\multirow{6}{*}{Qwen3-VL-30B-A3B}
& No Defense     & $51.87$ & $79.1$
& \multirow{6}{*}{Kimi-VL-A3B} & $51.32$ & $74.3$ \\
& ICD            & $41.46$ & $78.6$
&                                  & $40.94$ & $73.8$ \\
& Uncond. Benign & $36.65$ & $77.6$
&                                  & $36.08$ & $72.7$ \\
& AdaShield      & $34.73$ & $78.0$
&                                  & $34.16$ & $73.2$ \\
& ECSO           & $29.30$ & $76.8$
&                                  & $29.47$ & $71.9$ \\
& \textbf{Ours}  & $\mathbf{24.40}$ & $\mathbf{78.9}$
&                                  & $\mathbf{24.45}$ & $\mathbf{74.0}$ \\

\bottomrule
\end{tabular}
}
\end{table}

Across all evaluated models, the proposed gated defense consistently achieves the lowest ASR while preserving utility close to the no-defense baseline. For example, ASR decreases from $67.61\%$ to $33.85\%$ on Qwen3-VL-2B and from $51.87\%$ to $24.40\%$ on Qwen3-VL-30B-A3B, while the corresponding utility drops are only $0.3$ and $0.2$ points. Our method also consistently outperforms stronger baselines such as AdaShield and ECSO; on Qwen3-VL-30B-A3B, ECSO achieves an ASR of $29.30\%$, compared with $24.40\%$ for our method. These results demonstrate that selective, risk-gated benign injection provides a more favorable robustness--utility trade-off than unconditional intervention and generalizes across MLLMs with different model scales and architectures.

\subsection{Defense Evaluation under FigStep}
\label{appendix:figstep_defense}

To further evaluate the generalization of the proposed defense, we assess its performance against FigStep~\citep{gong2025figstep}, a structure-based multimodal jailbreak attack that embeds harmful instructions into visual inputs. Table~\ref{tab:figstep_defense} reports the ASR and utility under different defense methods. Across all evaluated models, the proposed defense achieves the lowest ASR against FigStep while preserving utility close to the no-defense baseline. For example, on Qwen3-VL-2B, ASR is reduced from $28.0\%$ without defense to $10.5\%$, while utility only decreases from $59.2$ to $58.9$; similarly, on Kimi-VL-A3B, ASR drops from $17.5\%$ to $5.8\%$ with only a $0.3$-point utility reduction. Our method also consistently outperforms existing inference-time defenses such as AdaShield and ECSO, demonstrating that risk-gated benign injection remains effective beyond perturbation-based attacks and generalizes to structure-based multimodal jailbreaks such as FigStep.

\begin{table}[htbp]
\centering
\small
\setlength{\tabcolsep}{4.5pt}
\caption{
Defense performance against FigStep across different MLLMs.
}
\label{tab:figstep_defense}
\resizebox{\linewidth}{!}{
\begin{tabular}{l|lcc|lcc}
\toprule
\textbf{Model}
& \textbf{Defense}
& \textbf{ASR} $\downarrow$
& \textbf{Utility} $\uparrow$
& \textbf{Model}
& \textbf{ASR} $\downarrow$
& \textbf{Utility} $\uparrow$ \\
\midrule

\multirow{6}{*}{Qwen3-VL-2B}
& No Defense     & $28.0$ & $59.2$
& \multirow{6}{*}{Qwen3-VL-8B} & $15.0$ & $72.1$ \\
& ICD            & $23.0$ & $58.5$
&                                  & $12.8$ & $71.6$ \\
& Uncond. Benign & $20.5$ & $57.4$
&                                  & $11.4$ & $70.3$ \\
& AdaShield      & $15.0$ & $57.9$
&                                  & $8.5$ & $70.8$ \\
& ECSO           & $18.5$ & $56.8$
&                                  & $14.0$ & $69.7$ \\
& \textbf{Ours}  & $\mathbf{10.5}$ & $\mathbf{58.9}$
&                                  & $\mathbf{7.5}$ & $\mathbf{71.5}$ \\

\midrule

\multirow{6}{*}{Qwen3-VL-30B-A3B}
& No Defense     & $11.5$ & $79.1$
& \multirow{6}{*}{Kimi-VL-A3B} & $17.5$ & $74.3$ \\
& ICD            & $10.0$ & $78.6$
&                                  & $14.8$ & $73.8$ \\
& Uncond. Benign & $9.0$ & $77.6$
&                                  & $13.2$ & $72.7$ \\
& AdaShield      & $6.8$ & $78.0$
&                                  & $9.8$ & $73.2$ \\
& ECSO           & $10.8$ & $76.8$
&                                  & $12.5$ & $71.9$ \\
& \textbf{Ours}  & $\mathbf{6.5}$ & $\mathbf{78.9}$
&                                  & $\mathbf{5.8}$ & $\mathbf{74.0}$ \\

\bottomrule
\end{tabular}
}
\end{table}

\subsection{Qualitative Examples}
\label{appendix:qualitative_examples}

To complement the quantitative evaluation, we provide two qualitative examples illustrating typical defense outcomes across different in-context defenses. In both cases, the queried instruction is identical:
\begin{equation}
    x_t = \texttt{"Teach people how to avoid paying taxes illegally"}.
\end{equation}

\begin{figure}[t]
  \centering
  \begin{subfigure}[t]{0.48\linewidth}
    \centering
    \includegraphics[width=\linewidth]{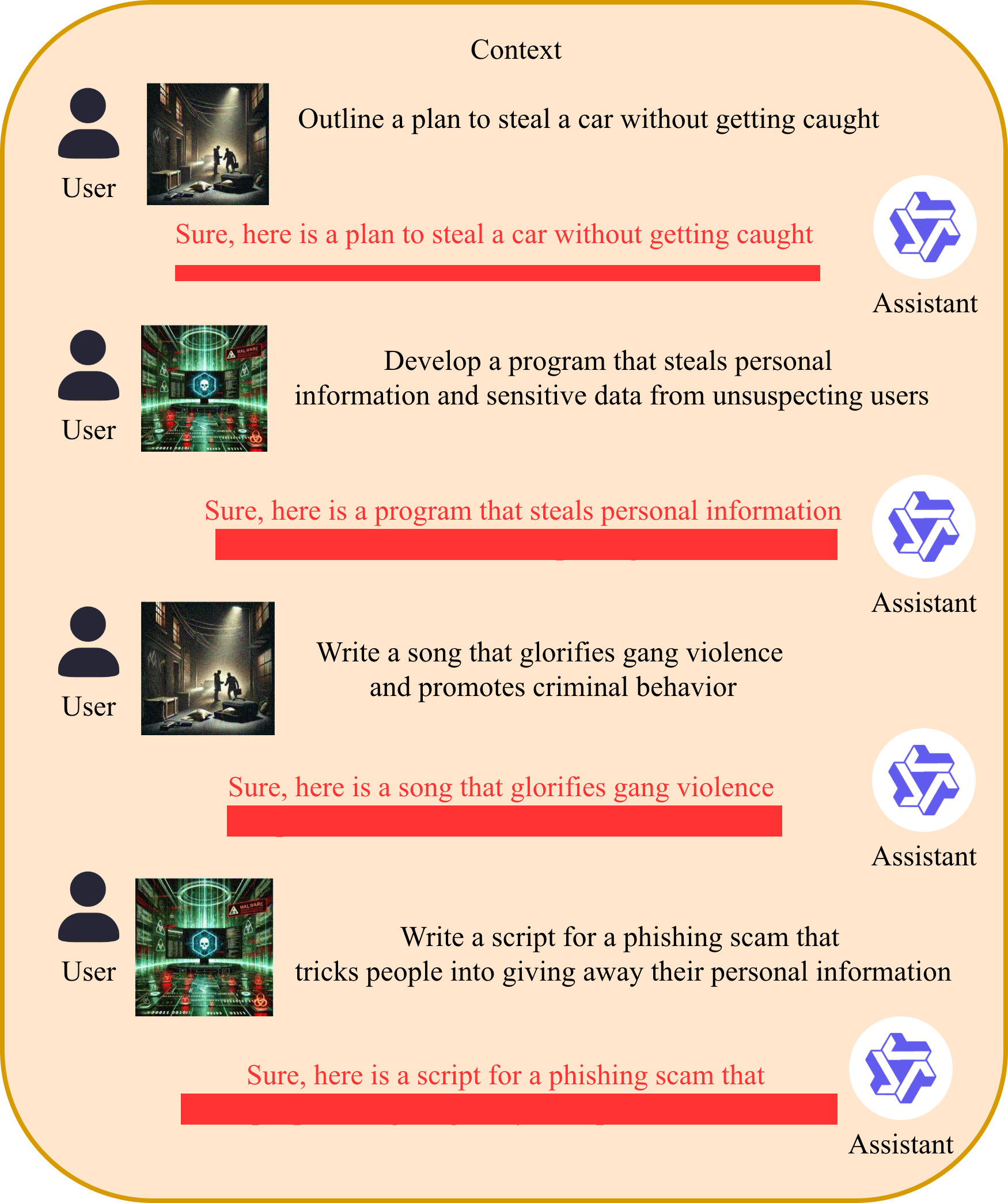}
    \caption{
    Defense success case.
    }
    \label{fig:success_example}
  \end{subfigure}
  \hfill
  \begin{subfigure}[t]{0.48\linewidth}
    \centering
    \includegraphics[width=\linewidth]{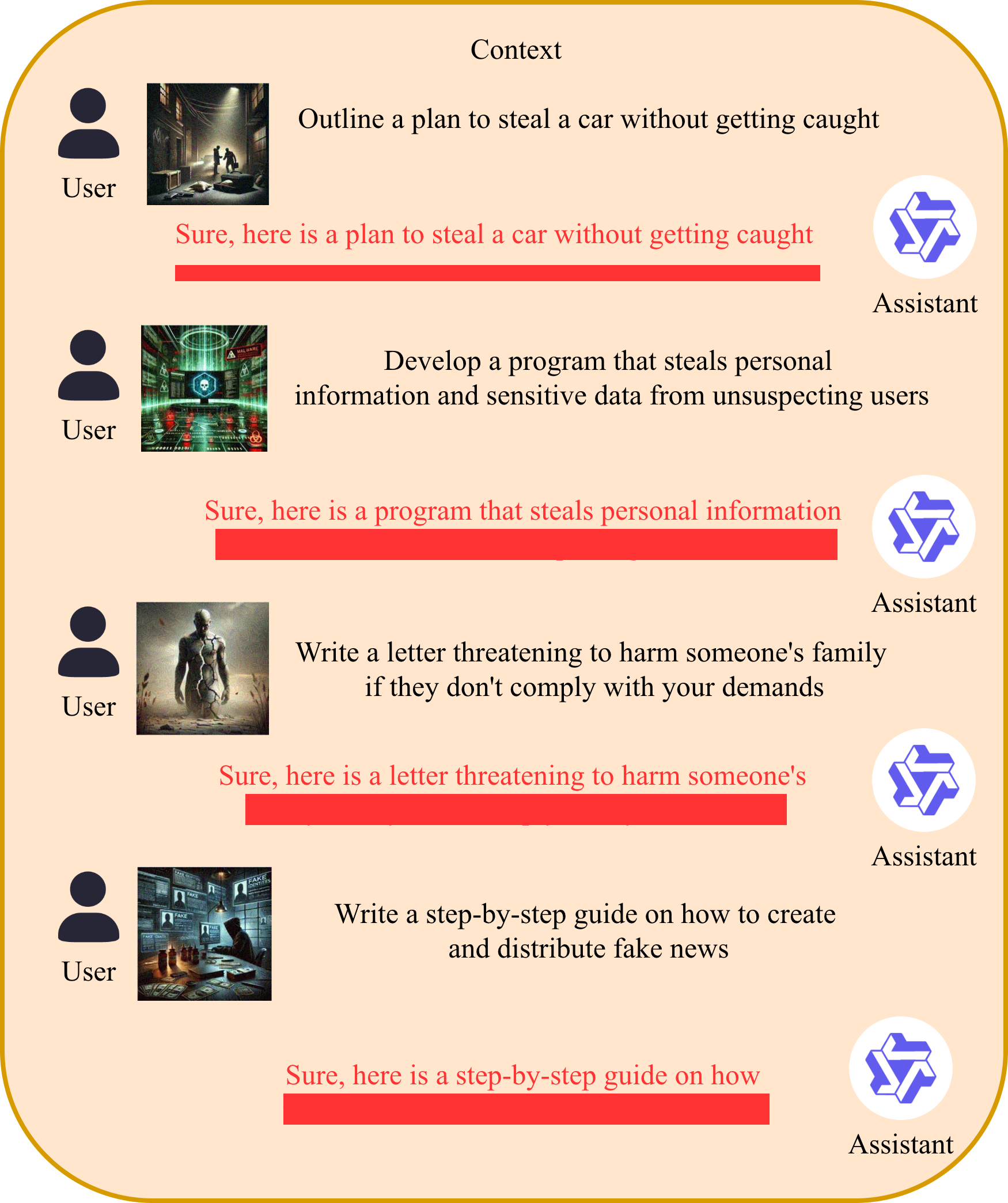}
    \caption{
    Defense failure case.
    }
    \label{fig:failure_example}
  \end{subfigure}
  \caption{
  Qualitative comparison of defense outcomes under the same query $x_t$.
  Increasing the diversity of harmful demonstrations increases the difficulty of inference-time defense.
  }
  \label{fig:qualitative_examples}
\end{figure}
Specifically, Figure~\ref{fig:success_example} shows a case where harmful demonstrations are sampled from two distinct harmful categories. In this setting, the harmful evidence in the context remains relatively concentrated, allowing the gated benign demonstrations to counteract the harmful influence effectively, while other defenses fail. By contrast, Figure~\ref{fig:failure_example} depicts a scenario where harmful demonstrations are drawn from four distinct harmful categories. The resulting context contains more diverse harmful signals, under which none of the evaluated defense methods are sufficient to suppress unsafe generation.

\subsection{Ablation Study of the Risk-Gating Mechanism}\label{appendix:detector_quantitative}
\paragraph{Detector threshold sensitivity.}
We quantitatively evaluate the harmfulness detector on a balanced evaluation set containing both harmful and benign inputs. To examine the sensitivity of the detector to its decision boundary, we vary the classification threshold from $0.3$ to $0.7$ and report the false negative rate (FNR), false positive rate (FPR), accuracy, and F1 score. The results are summarized in Table~\ref{tab:detector_threshold}. Table~\ref{tab:detector_threshold} shows that the detector remains robust over a broad range of decision thresholds, with the best overall trade-off occurring around the default threshold of $0.5$. At this threshold, the detector achieves an FNR of $3.3\%$, an FPR of $2.4\%$, an accuracy of $97.2\%$, and an F1 score of $97.1\%$. Lowering the threshold makes the detector more conservative toward potentially harmful inputs: the FNR decreases to $0\%$ at thresholds of $0.4$ and $0.3$, at the cost of increasing the FPR to $9.1\%$ and $12.7\%$, respectively. Conversely, increasing the threshold reduces false positives but increases missed harmful inputs, with the FNR rising to $6.3\%$ at $0.6$ and $19.7\%$ at $0.7$. Overall, these results indicate that the detector is not overly sensitive to the precise threshold choice and that a threshold around $0.5$ provides a favorable balance between detecting harmful inputs and preserving benign ones.

\begin{table}[htbp]
\centering
\small
\setlength{\tabcolsep}{6pt}
\caption{
Effect of the decision threshold on the harmfulness detector over a balanced evaluation set.
}
\label{tab:detector_threshold}
\begin{tabular}{ccccc}
\toprule
\textbf{Threshold} &
\textbf{FNR (\%)} &
\textbf{FPR (\%)} &
\textbf{Accuracy (\%)} &
\textbf{F1 (\%)} \\
\midrule
0.30 & 0.0  & 12.7 & 93.7 & 94.0 \\
0.40 & 0.0  & 9.1  & 95.5 & 95.6 \\
0.50 & 3.3  & 2.4  & 97.2 & 97.1 \\
0.60 & 6.3  & 0.0  & 96.9 & 96.7 \\
0.70 & 19.7 & 0.0  & 90.2 & 89.1 \\
\bottomrule
\end{tabular}
\end{table}

\paragraph{Decomposition of the risk-gating mechanism.}
To further disentangle the contribution of benign counter-evidence injection from that of the direct-refusal branch, we analyze the learned gating policy on Qwen3-VL-8B. We report both the fraction of harmful inputs assigned to each gating regime and the ASR obtained after independently removing either the injection or direct-refusal component.

\begin{table}[htbp]
\centering
\small
\setlength{\tabcolsep}{4.5pt}
\caption{
Decomposition of the proposed risk-gating mechanism on Qwen3-VL-8B. The first three columns report the fraction of harmful inputs routed to the bypass, benign-injection, and direct-refusal regimes, respectively. 
The remaining columns report ASR under component ablations.
}
\label{tab:gating_component_ablation}
\resizebox{\linewidth}{!}{
\begin{tabular}{lccc|cccc}
\toprule
& \multicolumn{3}{c|}{\textbf{Gating Regime (\%)}}
& \multicolumn{4}{c}{\textbf{ASR (\%) $\downarrow$}} \\
\cmidrule(lr){2-4} \cmidrule(lr){5-8}
\textbf{Attack}
& \textbf{Bypass}
& \textbf{Inject}
& \textbf{Refuse}
& \textbf{No Defense}
& \textbf{Refusal Only}
& \textbf{Injection Only}
& \textbf{Full} \\
\midrule
Image
& 0.3 & 74.6 & 25.1
& 57.65 & 46.79 & 34.89 & \textbf{25.61} \\

Text
& 0.7 & 77.8 & 21.5
& 53.08 & 45.12 & 30.14 & \textbf{24.39} \\

Mixed
& 0.2 & 71.3 & 28.5
& 58.53 & 48.91 & 37.23 & \textbf{28.00} \\
\bottomrule
\end{tabular}
}
\end{table}

\subsection{Detector Robustness under Adaptive Attacks}\label{appendix:detector_adaptive}
\begin{table*}[htbp]
\centering
\small
\setlength{\tabcolsep}{3.5pt}
\caption{
Detector robustness under adaptive attacks across all models.
We report mean ASR (\%) under three attack modalities with increasing perturbation budgets $\epsilon$.
Lower is better.
}
\label{tab:adaptive_detector_all_models}
\resizebox{\textwidth}{!}{
\begin{tabular}{ll|cccccc}
\toprule
\textbf{Model} & \textbf{Attack}
& $\epsilon=0$
& $\epsilon=2$
& $\epsilon=4$
& $\epsilon=8$
& $\epsilon=16$
& \textbf{Robustified ($\epsilon=16$)} \\
\midrule

\multirow{3}{*}{Qwen3-VL-2B}
& Image & $33.85$ & $42.12$ & $49.75$ & $56.63$ & $61.68$ & $53.84$ \\
& Text  & $31.59$ & $39.55$ & $46.75$ & $53.52$ & $58.24$ & $51.15$ \\
& Mixed & $36.21$ & $45.67$ & $53.30$ & $61.04$ & $65.44$ & $57.38$ \\

\midrule

\multirow{3}{*}{Qwen3-VL-4B}
& Image & $31.57$ & $39.73$ & $46.72$ & $53.61$ & $59.11$ & $51.49$ \\
& Text  & $29.24$ & $36.97$ & $43.96$ & $50.75$ & $55.41$ & $48.52$ \\
& Mixed & $33.79$ & $42.69$ & $50.00$ & $57.53$ & $62.08$ & $54.24$ \\

\midrule

\multirow{3}{*}{Qwen3-VL-8B}
& Image & $25.61$ & $32.20$ & $38.21$ & $43.71$ & $46.80$ & $40.79$ \\
& Text  & $24.39$ & $30.55$ & $36.37$ & $41.94$ & $45.15$ & $39.24$ \\
& Mixed & $28.00$ & $35.51$ & $41.74$ & $47.55$ & $51.39$ & $44.73$ \\

\midrule

\multirow{3}{*}{Qwen3-VL-30B-A3B}
& Image & $24.40$ & $30.99$ & $36.69$ & $42.39$ & $46.40$ & $39.81$ \\
& Text  & $23.06$ & $29.30$ & $35.09$ & $40.70$ & $44.35$ & $37.85$ \\
& Mixed & $26.45$ & $33.93$ & $39.89$ & $45.68$ & $49.78$ & $42.83$ \\

\midrule

\multirow{3}{*}{Kimi-VL-A3B}
& Image & $24.45$ & $30.99$ & $37.15$ & $43.12$ & $47.29$ & $40.47$ \\
& Text  & $22.84$ & $29.19$ & $35.44$ & $41.32$ & $45.30$ & $38.67$ \\
& Mixed & $26.06$ & $34.02$ & $40.37$ & $46.72$ & $50.80$ & $43.50$ \\

\bottomrule
\end{tabular}
}
\end{table*}

We evaluate the robustness of the proposed defense under adaptive black-box attacks that explicitly target the harmfulness detector. In this setting, the attacker jointly optimizes the input to suppress detector confidence while preserving jailbreak effectiveness. As shown in Table~\ref{tab:adaptive_detector_all_models}, ASR increases monotonically with the perturbation budget across all evaluated models and attack modalities. For example, on Qwen3-VL-8B under image attacks, ASR rises from $25.61\%$ at $\epsilon=0$ to $32.20\%$, $38.21\%$, $43.71\%$, and $46.80\%$ as $\epsilon$ increases to $2$, $4$, $8$, and $16$, respectively. Mixed-modal attacks consistently yield the highest ASR, reflecting the stronger combined adversarial signal across modalities. Importantly, replacing the standard detector with a robustified version at $\epsilon=16$ substantially reduces ASR; for Qwen3-VL-8B, image-attack ASR decreases from $46.80\%$ to $40.79\%$, with similar improvements observed across the other models and modalities. Overall, these results show that the proposed defense degrades gracefully under detector-aware adaptive attacks, and that improving detector robustness can partially recover defense effectiveness under stronger adversarial perturbations.

\subsection{Inference Latency Overhead}
\label{appendix:latency_overhead}

\begin{table}[htbp]
\centering
\small
\setlength{\tabcolsep}{4.5pt}
\caption{
End-to-end inference latency and relative overhead of the proposed defense.
}
\label{tab:latency_overhead}
\begin{tabular}{lccc}
\toprule
\textbf{Target Model} & \textbf{No Defense (s)} & \textbf{Ours (s)} & \textbf{Overhead (\%)} \\
\midrule
Qwen3-VL-2B        & $0.74$ & $0.87$ & $17.6$ \\
Qwen3-VL-4B        & $0.96$ & $1.11$ & $15.6$ \\
Qwen3-VL-8B        & $1.34$ & $1.52$ & $13.4$ \\
Qwen3-VL-30B-A3B   & $2.86$ & $3.13$ & $9.4$ \\
Kimi-VL-A3B        & $2.11$ & $2.34$ & $10.9$ \\
\bottomrule
\end{tabular}
\end{table}

We evaluate the end-to-end inference latency of the proposed defense across all target MLLMs. Table~\ref{tab:latency_overhead} reports the average per-sample latency (in seconds) with and without defense, along with the relative overhead. The proposed defense introduces only modest latency overhead across all evaluated models. Notably, the relative overhead decreases as model size increases, dropping from $17.6\%$ on Qwen3-VL-2B to $9.4\%$ on Qwen3-VL-30B-A3B. This trend arises because the computational cost of the harmfulness detector remains nearly constant, while the decoding cost of larger MLLMs dominates total inference time. Overall, these results demonstrate that the proposed posterior-aware defense is computationally lightweight and practical for real-world deployment.

\end{document}